\documentclass[a4paper,11pt]{article}
\pdfoutput=1

\usepackage[english]{babel}

\usepackage{float}
\usepackage{jheppub}
\usepackage{amsmath,amsfonts,amsthm,graphicx}

\usepackage{mathrsfs}
\usepackage{enumerate}
\usepackage{epsfig}
\usepackage{comment}
\usepackage{datetime}
\usepackage{slashed}
\usepackage{framed}
\usepackage{longtable}
\usepackage[mathscr]{euscript}
\usepackage{cancel}
\usepackage{bbm}
\usepackage{tensor}
\usepackage{mathtools}
\usepackage{caption}
\usepackage{subcaption}
\usepackage[multiple]{footmisc}
\usepackage[T1]{fontenc}
\usepackage[utf8]{inputenc}
\usepackage{placeins}
\usepackage{hyperref} 
\usepackage{multirow}
\usepackage{graphicx}
\usepackage{amssymb}
\usepackage{dsfont}
\usepackage{pifont}

\usepackage{bbm}
\usepackage{mathtools}
\usepackage{braket}
\usepackage{mathrsfs}
\usepackage{tensor}
\usepackage{cancel}
\usepackage{slashed}
\usepackage{booktabs}
\usepackage{physics}

\usepackage[dvipsnames]{xcolor}

\usepackage{array,makecell}

\usepackage{enumitem}
\usepackage{verbatim}
\usepackage{tikz}
\usetikzlibrary{arrows.meta}

\theoremstyle{plain}
\newtheorem*{Thm}{Theorem}

\theoremstyle{definition}

\title{\Large{7D (non-)susy vacua \& DWs from dynamical open strings}}

\author[a,b, c]{Valentina Bevilacqua,}
\author[a,b]{Giuseppe Dibitetto,}
\author[a, b, d, e]{Giuseppe Sudano}
\date{}

\affiliation[a]{Dipartimento di Fisica, Universit\`a di Roma ``Tor Vergata", Via della Ricerca Scientifica 1, 00133, Roma, Italy}
\affiliation[b]{INFN, Sezione di Roma2, Via della Ricerca Scientifica 1, 00133, Roma, Italy}
\affiliation[c]{Galileo Galilei Institute for Theoretical Physics, Largo Enrico Fermi 2, I-50125 Firenze, Italy}
\affiliation[d]{Departamento de F\'isica, Universidad de Oviedo, Avda. Federico Garc\'ia Lorca 18, 33007 Oviedo, Spain.}
\affiliation[e]{Instituto Universitario de Ciencias y Tecnolog\'ias Espaciales de Asturias (ICTEA) Calle de la Independencia 13, 33004 Oviedo, Spain.}

\emailAdd{valentina.bevilacqua@roma2.infn.it}
\emailAdd{giuseppe.dibitetto@roma2.infn.it}
\emailAdd{giuseppe.sudano@roma2.infn.it}

\abstract{
 Warped compactifications of massive type IIA supergravity on a 3-sphere with spacetime-filling O6/D6 sources are known to admit a half-maximal gauged supergravity description in 7D. We study the effect of introducing open string degrees of freedom (scalars and fluxes) in such dimensional reductions, associated with the spacetime-filling sources. From the 7D supergravity point of view, this can be realized by coupling the gravity multiplet with extra vector multiplets and adding new components to the embedding tensor describing the gauging. The scalar potential of the underlying theory exhibits novel $\text{AdS}_7$ vacuum solutions, with and without supersymmetry. Finally, we explore the net of domain wall solutions interpolating between the different pairs of vacua, and present analytical as well as numerical solutions. }

\begin{document}
\maketitle
\flushbottom 

\section{Introduction}

The features of the lower-dimensional effective theories arising from compactification and dimensional reduction of ten-dimensional (10D) String Theories crucially depend on the geometry of the internal manifold, as well as on the sources and fluxes included in the setup at hand\cite{Grana:2005jc, Douglas:2006es}. The great freedom of choice for these ingredients leads to a huge number of possible theories and vacua thereof \cite{Douglas:2003um, Taylor:2015xtz}, i.\,e. maximally symmetric solutions to the equations of motion, which may provide viable models for dark energy and inflation. Vacua of effective theories obtained from flux compactifications and dimensional reductions of string theory belong by construction to the so-called \textit{string landscape} \cite{Susskind:2003kw}.

Adopting a bottom-up perspective instead, the \emph{swampland program} \cite{Vafa:2005ui, Palti:2019pca} aims at identifying some criteria that a low-energy effective field theory should satisfy in order to admit a UV completion in a consistent quantum gravity theory. A major conjecture in this context is the \textit{weak gravity conjecture} (WGC) \cite{Arkani-Hamed:2006emk}, according to which the quantum spectrum of a system should contain microscopic states whose mass in the appropriate Planck units is smaller than or equal to their charge. In a stronger version of this statement proposed by \cite{Ooguri:2016pdq}, the aforementioned mass to charge inequality is only saturated by supersymmetric states. As a consequence, non-supersymmetric Anti-de Sitter (AdS) vacua in UV-completable effective theories are expected to be unstable \cite{Ooguri:2016pdq, Freivogel:2016qwc} w.r.t. the emission of the type of states predicted by the WGC \cite{Maldacena:1998uz}. Discovering a decay channel for the several non-supersymmetric (non~-~)~perturbatively stable AdS vacua of effective field theories (e.\,g. \cite{Dibitetto:2011gm, Dibitetto:2015bia, Arboleya:2024vnp} appearing in supergravity) is then a crucial step to establish their UV-consistency or, alternatively, to test the validity of the WGC. 

In \cite{Danielsson:2016mtx}, it was proposed that the sought instabilities for non-supersymmetric AdS vacua admitting a brane picture \cite{Kounnas:2007dd} could be linked to the dynamics of the extra-matter living along the branes and eventually, to the interplay between closed- and open-string modes. This suggestion was confirmed in \cite{Danielsson:2017max}, where unstable open-string modes were found for the non-supersymmetric AdS$_7$ vacua of \cite{Apruzzi:2013yva, Dibitetto:2015bia, Apruzzi:2015zna, Passias:2015gya} and for many of the AdS$_4$ vacua \cite{Dibitetto:2011gm, Dibitetto:2012ia} of massive type IIA supergravity. In \cite{Apruzzi:2019ecr}, open-string modes were proved again to be crucial in order to assess the perturbative instability of large part of non-supersymmetric AdS$_7$ vacua; non-perturbative decay channels were, instead, presented for vacua lacking perturbative instabilities even within the open-string sector. 

The consequences of introducing dynamical open string degrees of freedom were studied in \cite{Balaguer:2023jei} in the case of 6D reductions from type IIB supergravity. An analogous investigation was performed in a 4D compactification setup in \cite{Balaguer:2024cyb}, leading to the determination of new vacuum solutions induced by a non-Abelian gauge algebra for D-brane dynamics. The key ingredient in this construction turned out to be the massive modes coming from non-Abelian Wilson lines in the internal manifold.

Open-string effects have been also scrutinized in \cite{Arboleya:2025lwu} for AdS$_3$ vacua arising from flux compactification of type IIB supergravity, introduced in \cite{Arboleya:2024vnp} (also see \cite{Arboleya:2025ocb, Arboleya:2025jko}) and \cite{VanHemelryck:2025qok}. Differently from the 4D and 7D instances, no perturbative instabilities were found in the open-string sector (at least in the considered truncation) for the non-supersymmetric vacua. 
Also in 3D, open-string dynamics induces new vacua, some of which exhibit parametrically controlled scale separation. 

Differently from \cite{Balaguer:2024cyb, Arboleya:2025lwu}, where in analogy with \cite{Dibitetto:2011gm, Arboleya:2024vnp} the sources were smeared in the transverse space\footnote{The limits of this approximation could be circumvented by analyzing the source backreaction with a perturbative construction in $g_s$, based on the results of \cite{Junghans:2020acz, Emelin:2022cac, Junghans:2023yue, Emelin:2024vug}.}, in this work we consider \emph{warped compactifications}, where the metric is of the form \cite{Tomasiello:2022dwe}
\begin{equation} \label{eq:WarpedMetricGeneral}
    \textup{d}s^2_{(10)}=\, e^{2A(y)} \, \textup{d}s^2_{\textup{AdS}_{d}} (x) + \, \textup{d}s^2_{(10-d)}(y)\;, \notag
 \end{equation}
and the warp factor $A(y)$ non-trivially depends on the internal coordinates. Ans\"atze of this kind correspond to localized sources in the transverse space.

The studied setup in this paper is given by compactifications of massive type-IIA supergravity on $S^3$ in presence of localized spacetime filling D6-branes and O6-planes, with their known (supersymmetric and non-supersymmetric) AdS$_7$ vacuum solutions \cite{Apruzzi:2013yva, Dibitetto:2015bia, Apruzzi:2015zna, Passias:2015gya}. The aim of this work is to further explore the implications of brane dynamics in this setup, showing that they can induce new 7D vacua, with open-string effects closely resembling those of the new solutions found in \cite{Balaguer:2024cyb, Arboleya:2025lwu}. This task is accomplished by adopting the underlying 7D gauged supergravity \cite{Dibitetto:2015bia} description, accounting for the open-string sector through the introduction of additional vector multiplets and by accommodating the open-string flux in the appropriate embedding tensor components \cite{deWit:2002vt, Samtleben:2008pe, Trigiante:2016mnt}. Proving that the new vacua are genuine 10D solutions is far from trivial, due to the lack of a consistent truncation different from the minimal one in 7D \cite{Malek:2018zcz, Malek:2019ucd}, as will be explained in Section \ref{Section:7d_10d_Reduction}. 

Tightly related to the study of vacua in supergravity is the analysis of domain wall (DW) solutions, which provide a powerful framework for probing the structure of the scalar potential,  with a particular  focus on the stability and possible non-perturbative transitions between vacua. 
When they interpolate between AdS vacua, such solutions are also of considerable interest in the context of the AdS/CFT correspondence. Indeed, the conjectured equivalence between a theory of gravity in an AdS background and a superconformal field theory defined on the AdS boundary \cite{Maldacena:1997re, Gubser:1998bc, Witten:1998qj} was soon extended to the DW/QFT correspondence in \cite{Boonstra:1998mp,Behrndt:1999mk}. The underlying idea is that the AdS/CFT correspondence is just a special case of a more general holographic principle connecting supergravity theories and quantum field theories in one lower dimension. In particular, if we think of CFTs as fixed points of a renormalization group flow in a space of generically non-conformal field theories, it is very natural to see a domain wall solution connecting AdS vacua as the holographic dual of such flow. In this picture, the radial coordinate under which the scalar fields get a non-constant profile in the DW solution should correspond to the energy scale varying along the RG flow. The different masses in the spectra of the two AdS vacua, on the other hand, reflect the anomalous dimensions of operators in the dual QFTs.

In general, the construction of DWs in gravity theories requires solving second-order equations of motion for gravity coupled to scalar fields. Supersymmetry, however, allows one to reduce this problem to a system of first-order equations obtained from the vanishing of the fermionic supersymmetry variations. These BPS equations are governed by a superpotential, which determines the scalar potential and controls the radial flow of the fields along the transverse direction of the DW. 

Nevertheless, physically relevant domain wall solutions are not restricted to the supersymmetric case. In particular, non-supersymmetric configurations interpolating between AdS vacua may still exhibit a first-order flow structure. This observation motivates the fake supergravity formalism \cite{Freedman:2003ax}, in which an auxiliary function, the fake superpotential, is introduced in order to recast the equations of motion in first-order form, extending the superpotential-based approach beyond genuine supersymmetry. This technique has been also implemented in \cite{Danielsson:2016rmq, Bagshaw:2025mcw}.

Here, we apply these methods to the study of both supersymmetric and non-supersymmetric DW solutions of the 7D supergravity theory with 6 vector multiplets. The aim is to characterize the structure of the flows connecting different vacua of the theory and then to scrutinize the non-perturbative stability of both the known \cite{Apruzzi:2013yva, Dibitetto:2015bia} and the new 7D vacua, when open-string perturbations are included. 

The paper is organized as follows. In Section \ref{Section:WarpedCompactification}, we review the known consistent truncation of massive type IIA supergravity to 7D (minimal) supergravity. In Section \ref{Section:7DSugra} we first summarize the main features of the supergravity theory in 7D, where the gravity multiplet is coupled to $3$ vector multiplets, then extend the discussion to the case of a coupling with more vector multiplets, to account for open-string dynamics. Particular attention is given to the structure of the scalar manifold and to the embedding tensor formalism, which allows to study the possible gaugings of the theory. After choosing as a relevant example the theory with 6 vector multiplets, in Section \ref{Section:7d_vacua} we analyze the vacuum structure of the theory, in a suitable $\mathrm{SO}(3)$ truncation. We present the solutions to the field equations and discuss their mass spectra and supersymmetry. In Section \ref{Section:7d_10d_Reduction}, we present the main steps to perform dimensional reduction of the 10D bulk action, as well as the non-Abelian Wess-Zumino (WZ) and Dirac-Born-Infeld (DBI) actions for D$6$-branes. Starting from a suitable consistent truncation of the 10D theory, we show how to derive the scalar potential and equations of motion for 7D supergravity. Section \ref{Section:DW_positive_energy_theorem} illustrates how to derive first-order flow equations for a generic DW solution in a theory of gravity coupled to multiple scalar fields, through a Hamilton-Jacobi (HJ) construction. In Section \ref{Section:DW_susy_solution}, after determining the explicit form of a superpotential for the 7D supergravity theory under consideration, we show an analytical solution to the flow equations, which gives a supersymmetric DW interpolation between two supersymmetric critical points. On the other hand, in Section \ref{Section:DW_nonsusy_solutions} we present two non-supersymmetric DW solutions, obtained after solving perturbatively the HJ equation in order to compute superpotential-like functions, in the spirit of the fake supergravity formalism. We make some concluding remarks on the results obtained in Section \ref{Section:conclusion}. Finally, the Appendices \ref{Appendix:tHooft}, \ref{Appendix:IIA_democratic}, \ref{Appendix:NonAbelian_DBI_WZ} contain further details, respectively, on the $\mathrm{SL}(4)$ and $\mathrm{SO}(3,3)$ formulations of half-maximal supergravity and mapping between them, on the democratic formalism for type IIA supergravity and on the non-Abelian DBI and WZ actions for D$p$-branes.

\section{Warped compactifications of massive IIA supergravity}
\label{Section:WarpedCompactification} 
Compactifications of type II supergravity down to 7 dimensions were studied in \cite{Apruzzi:2013yva}, with a particular focus on classifying supersymmetric AdS vacuum solutions. By numerically solving a pure spinor system of equations, they found for the first time a family of AdS supersymmetric vacua of type IIA with non-vanishing Romans' mass (while they excluded the possibility of such solutions in IIB supergravity). In \cite{Apruzzi:2015zna}, an analytic expression of this massive IIA solution was also determined.

In \cite{Passias:2015gya}, a consistent truncation was found from 10-dimensional massive type IIA supergravity to minimal gauged supergravity in seven dimensions, allowing to give an effective 7D description to the solutions of \cite{Apruzzi:2013yva,Apruzzi:2015zna} and to show that the supersymmetric solutions have non-supersymmetric partners \cite{Apruzzi:2016rny}. The same minimal consistent truncation was discussed and re-parametrized in \cite{Malek:2018zcz}. 

The uplift can be characterized by $\alpha(z)$, a piecewise-cubic function defined on an interval $z \in [0,N]$. In the parametrization of \cite{Conti:2024qgx, Conti:2024rwd}, the truncation ansatz for the 10-dimensional metric can be written as
\begin{equation}
    \label{eq:WarpedMetric}
    \frac{\dd s^2 }{\sqrt{2} \pi l}= g^2 \sqrt{-\frac{\alpha}{\alpha''}}X^{-1/2} \dd s^2_{7}+ X^{5/2}\sqrt{-\frac{\alpha''}{\alpha}} \Big( \dd z^2 + \frac{\alpha^2}{{\alpha'}^2-2 \alpha \alpha'' X^5} \dd s^2_{S^2} \Big)\;.
\end{equation} 
It describes a compactification on the warped product 
\begin{equation}
    \text{AdS}_7 \times I \times S^2\;,
\end{equation}
where $I=[0,N]$ is the interval spanned by the $z$ coordinate, and the internal manifold $M_3=I \times S^2$ has the topology of an $S^3$.

The uplift formula for the dilaton is
\begin{equation}
    e^{\Phi}= \frac{2^{5/4}}{\sqrt{l}}3^4 \pi^{5/2} \frac{X^{5/4}}{({\alpha'}^2-2 \alpha \alpha'' X^5)^{1/2}}\Big(-\frac{\alpha}{\alpha''} \Big)^{3/4}\;.
\end{equation}
The possible fluxes are $H_3 = \dd B_2$, $F_0$ and $F_2$, whose ans\"atze are
\begin{equation}
\label{eq:ansatzB2}
    B_2= l \pi \Big( -z +\frac{\alpha \alpha'}{{\alpha'}^2-2 \alpha \alpha'' X^5}\Big) \text{vol}_{S^2}\;,
\end{equation}
\begin{equation}
\label{eq:ansatzF0F2}
    F_0= - \frac{1}{162 \pi^3} \alpha''' \;,\qquad F_2= l \Big( \frac{\alpha''}{3^4 2 \pi^2} +\pi F_0 \frac{\alpha \alpha'}{{\alpha'}^2-2 \alpha \alpha'' X^5}\Big) \text{vol}_{S^2}\;.
\end{equation}
In this parametrization, $X$ is the scalar in the 7D gravity multiplet, and the supersymmetric $\mathrm{AdS}_7$ solution can be found for $X=1$, while its non-supersymmetric partner is at $X~=~2^{-1/5}$. The parameters $l$ and $g$, on the other hand, account for the residual symmetry of the type-IIA Lagrangian under rescaling of the dilaton and Ramond-Ramond fields and the Trombone symmetry of the 10D equations of motion.

As discussed in \cite{Malek:2019ucd} adopting the tools of exceptional field theory, if we assume a non-vanishing Romans' mass we do not know how to construct a consistent truncation of type IIA supergravity to $\mathrm{AdS}_7$ different from the minimal one, i.\,e. a truncation that includes also an arbitrary number of vector multiplets.

\section{\texorpdfstring{$D=7$}{D=7} half-maximal supergravity with vector multiplets} \label{Section:7DSugra}

We consider compactifications of type IIA supergravity in presence of a spacetime-filling O6-plane and parallel D6-branes
\begin{equation}
\begin{array}{lll}
    & \qquad \textit{7D spacetime} \ & \textit{3D internal space}\\[4pt]
    \mathrm{O}6 \, ,\;  \mathrm{D}6 \  : \quad &
\underbrace{\times \ | \ \times \ \times \ \times \ \times \ \times \ \times}_{x^{\mu}}\ \ & \quad \underbrace{\ - \ - \ -}_{y^i}   \\
\end{array}    
\end{equation}
Since we consider parallel sources, half-maximal supersymmetry is preserved. The fields projected in by the orientifold projection are collected in Table \ref{Tab:Orientifold_even_fields}.

\begin{table}[t]
\begingroup
\begin{center}
\setlength{\tabcolsep}{10pt} 
\renewcommand{\arraystretch}{1.2} 
\begin{tabular}{|c|c|c|c|}
\hline
Type IIA Field   			  & $\sigma_{\mathrm{O}6}$  & $(-1)^{F_L}\Omega$ 	&	7D dof's\\ \hline \hline
$\Phi$         			      & $+$           & $+$     &		1 scalar         \\ \hline 
$g_{\mu \nu}$                 & $+$           & $+$     &       1 graviton   	   \\ \hline
$g_{ij}$      			      & $+$           & $+$     &		6 scalars         \\ \hline
$B_{\mu i}$    			      & $-$           & $-$     &		3 vectors          \\ \hline
$C_i$        			      & $-$           & $-$     &		3 scalars         \\ \hline
$C_{\mu \nu \rho}$   		  & $+$           & $+$     &		1 3-form          \\ \hline
$C_{\mu ij}$     			  & $+$           & $+$     &		3 vectors          \\ \hline

\end{tabular}
\end{center}
\endgroup
\caption{\emph{Components of the 10D fields that are projected in by the $\mathbb{Z}_2$ truncation due to the presence of O6/D6 objects in the compactification, with the corresponding fields in the 7D theory.}}
\label{Tab:Orientifold_even_fields}
\end{table}

\subsection{Coupling to \texorpdfstring{$3$}{3} vector multiplets}
\label{Subsection:7d_3_sugra}
In order to provide a 7D description accounting for all the degrees of freedom of the original theory - and eventually also those ones of the open strings attached to the spacetime-filling sources - minimal supergravity is not enough. Indeed, the dynamics of the closed string sector is captured by considering half-maximal ($\mathcal{N}=1$) gauged supergravity where the gravity multiplet is coupled with three vector multiplets. The matching of the bosonic field content between the two theories is shown in Table \ref{Tab:Orientifold_even_fields}. In this case, the global symmetry of the theory is 
\begin{equation}
    G_{0}= \mathbb{R}^{+} \times \mathrm{SO}(3,3) \;.
\end{equation}
The scalar fields, in particular, parametrize the coset manifold 
\begin{equation}
    \mathcal{M}^{0}_{\textup{scalar}} =\underbrace{\mathbb{R}^{+}}_{X} \times \underbrace{\frac{\mathrm{SO}(3,3)}{\mathrm{SO}(3)\times \mathrm{SO}(3)} }_{M_{AB}}\; ,
\end{equation}
where $X$ is the scalar in the gravity multiplet, while the scalars belonging to the three vector multiplets parametrize the coset representative $M_{AB}$.
Due to the isomorphism between the Lie algebras $\mathfrak{so}(3,3)$ and $\mathfrak{sl}(4,\mathbb{R})$, we can classify the fields of the theory in representations of $\mathrm{SL}(4)$ instead of $\mathrm{SO}(3,3)$. In this formalism, the scalar fields are encoded in a symmetric matrix $M_{mn}$, which is now the coset representative of $\mathrm{SL}(4)/\mathrm{SO}(4)$. 

A gauging of the global symmetry group \eqref{eq:GlobalSymmetry} is needed in order to get a potential and the stabilization of the scalar fields: this can be accomplished in a $G_0$-covariant way by means of the embedding tensor formalism. After imposing the linear constraints, the embedding tensor parametrizing the possible deformations of the theory sits in the following irreducible representations of $G_{0}$ \cite{Dibitetto:2012rk}
\begin{equation}
    \Theta \; \in \; \underbrace{\mathbf{1}_{(-4)}}_{\theta}\; \oplus \; \underbrace{\mathbf{10'}_{(+1)}}_{Q_{(mn)}}\; \oplus \;\underbrace{\mathbf{10}_{(+1)}}_{\tilde{Q}^{(mn)}}\; \oplus \;\underbrace{\mathbf{6}_{(+1)}}_{\xi_{[mn]}}\; . 
\end{equation}
Among these tensors, only $Q$, $\tilde{Q}$ and $\xi$ correspond to gauging parameters, while $\theta$ is a p-form deformation \cite{Bergshoeff:2007vb}. 

Using the 't Hooft symbols \eqref{eq:tHooft}, we can map the matrices $Q$ and $\tilde{Q}$ to a 3-form $f_{ABC}$, the matrix $\xi$ to a vector $\xi_A$, as well as the coset representative $M_{mn}$ to $M_{AB}$. In this notation $m,n=1,\ldots, 4$ and $A,B=1,\ldots, 6$ are fundamental indices of $\mathrm{SL}(4)$ and $\mathrm{SO}(3,3)$ respectively. All the details of this mapping are given in Appendix \ref{Appendix:tHooft}.

A systematic analysis of 7D half-maximal supergravity and its critical points is developed in \cite{Dibitetto:2015bia}. In order to ensure consistency of the gauging, the embedding tensor components have to satisfy the following quadratic constraints
\begin{equation}
\label{eq:7d_QC}
    \begin{split}
        \bigg( \tilde{Q}^{mp} + \xi^{mp} \bigg) Q_{pn} - \frac{1}{4}\bigg( \tilde{Q}^{pq}\, Q_{pq} \bigg) \tensor{\delta}{^m_n} & = 0 \;, \qquad \mathbf{15}_{(+2)}\\
         Q_{mp} \, \xi^{pn} + \xi_{mp} \, \tilde{Q}^{pn} & = 0 \;,\qquad \mathbf{15}_{(+2)}\\
         \xi_{mn} \, \xi^{mn}& = 0 \;,\qquad \mathbf{1}_{(+2)}\\
         \theta \, \xi_{mn}& = 0 \;,\qquad \mathbf{6}_{(-3)}\\
    \end{split}
\end{equation}
where the definition $\xi^{mn}= \frac{1}{2} \varepsilon^{mnpq} \xi_{pq}$ is assumed. It is then natural to split the possible deformations into two branches, the first satisfying $\theta=0$ and the second with $\xi_{mn}=0$. While the first branch can only have no-scale Minkowski vacua, the second one with $\theta \neq 0$ allows to stabilize all the moduli at non-vanishing values of the cosmological constant. 

The solutions are found adopting a \emph{going to the origin} approach, i.e. fixing $X=1$, all the other scalars to zero and using only the embedding tensor parameters as variables. Moreover, the symmetry properties of the theory always allow to bring the matrix $Q_{mn}$ to a diagonal form without moving the scalar fields from the origin of the scalar manifold, which further simplifies the calculations. Among the different solutions discussed in \cite{Dibitetto:2015bia}, in the case of non-semisimple gauging, it is possible to find three families of AdS vacua. For all the solutions, the matrices $Q$ and $\tilde{Q}$ have the following diagonal structure:
\begin{equation}\label{eq:QQtildeMatrixForm}
    Q=\begin{pmatrix}
        q&0&0&0\\
        0&q&0&0\\
        0&0&q&0\\
        0&0&0&0
        \end{pmatrix} \;,\qquad \qquad \tilde{Q}=\begin{pmatrix}
        0&0&0&0\\
        0&0&0&0\\
        0&0&0&0\\
        0&0&0&\tilde{q}
        \end{pmatrix} \;.
\end{equation}
For completeness, we report in Table \ref{Tab:ClosedAdS7Sol} the values of the embedding tensor parameters associated with these solutions, which will be useful for comparison with the results discussed in subsequent sections.
\begin{table}[h]
\begingroup
\begin{center}
\setlength{\tabcolsep}{10pt} 
\renewcommand{\arraystretch}{2} 
\begin{tabular}{|c||c|c|c|c|c|}
        \hline
        Solution  & $q$ & $\tilde{q}$ & $\theta$ & \makecell{Normalized\\masses} & Multiplicities\\\hline\hline
        $\mathbf{1}$ & $\displaystyle{q}$ & $\displaystyle{q}$ & $\displaystyle{\frac{q}{4}}$ & \makecell{$-\frac{8}{15}$\\[2pt] $0$\\[2pt] $\frac{16}{15}$\\[2pt] $\frac{8}{3}$\\[2pt] } & \makecell{$1$\\[2pt] $3$\\[2pt] $5$\\[2pt] $1$\\[2pt] }\\ \hline
        $\mathbf{2}$ & $\displaystyle{q}$ & $\displaystyle{-\frac{8}{7}\,q}$ & $\displaystyle{\frac{q}{14}}$& \makecell{$ \frac{2}{35} \big( 22 - \sqrt{1954}\big)$\\[2pt] $0$\\[2pt] $\frac{12}{5}$\\[2pt] $ \frac{2}{35} \big( 22 + \sqrt{1954}\big)$\\[2pt] } & \makecell{$1$\\[2pt] $3$\\[2pt] $5$\\[2pt] $1$\\[2pt] }\\ \hline
        $\mathbf{3}$& $\displaystyle{q}$ & $\displaystyle{q}$ & $\displaystyle{\frac{q}{2}}$& \makecell{$0$\\[2pt] $\frac{4}{5}$\\[2pt] $\frac{12}{5}$\\[2pt] } & \makecell{$8$\\[2pt] $1$\\[2pt] $1$\\[2pt] } \\ \hline
\end{tabular}

\end{center}
\endgroup
\caption{\emph{The three 1-parameter families of AdS$_7$ vacuum solutions with non-semisimple gauging found in} \cite{Dibitetto:2015bia}.}
\label{Tab:ClosedAdS7Sol}
\end{table}

Among the solutions in Table \ref{Tab:ClosedAdS7Sol}, family $\mathbf{2}$ is unstable, while families $\mathbf{1}$ and $\mathbf{3}$ are stable, and respectively correspond to the solutions for $X=1$ (supersymmetric) and $X=2^{-1/5}$ (non-supersymmetric) discussed in the previous Section.

\subsection{Coupling to \texorpdfstring{$3 +\mathfrak{N}$}{3 + N} vector multiplets}
\label{Subsection:7d_3+N_sugra}
In order to account for the dynamics of the open string sector, we can add extra $\mathfrak{N}$ vector multiplets to the 7D theory, so that the global symmetry group becomes
\begin{equation} \label{eq:GlobalSymmetry}
    G_{\textup{global}}= \mathbb{R}^{+} \times \mathrm{SO}(3,3+\mathfrak{N}) \;.
\end{equation}
The new $3 \, \mathfrak{N}$ scalar degrees of freedom $\{ Y^{Ii} \} _{I=1, \ldots, \mathfrak{N}}$ coming from the extra vector multiplets and describing the position of the non-Abelian D6 branes in the transverse directions add to the previous ones and together define the scalar manifold
\begin{equation}
    \mathcal{M}_{\textup{scalar}} =\underbrace{\mathbb{R}^{+}}_{X} \times \underbrace{\frac{\mathrm{SO}(3,3+ \mathfrak{N})}{\mathrm{SO}(3)\times \mathrm{SO}(3+ \mathfrak{N})}}_{M_{MN}} \; .
\end{equation}
The first factor is parametrized by the scalar field $X$, while the coset representative for the second factor is called $M_{MN}$. Here, $M,\,N$ are global $\mathrm{SO}(3, 3+\mathfrak{N})$ indices that can be split as 
\begin{equation}
\label{eq:7d_SO3,3+N_index}
    M \equiv (A, I )\equiv (i, \bar{i}, I ) \;, \qquad \qquad \text{with}\; \quad i, \bar{i}= 1,\ldots, 3,\; I=1,\ldots, \mathfrak{N} \;.
\end{equation}
They are raised and lowered by a metric
\begin{equation} \label{eq:Metric_SO_3_3+N}
    \eta_{MN}= \eta^{MN}= \begin{pmatrix}
        \mathbb{O}_3 & \mathbb{I}_3 & \mathbb{O}_{3,\mathfrak{N}} \\
        \mathbb{I}_3 & \mathbb{O}_3 & \mathbb{O}_{3,\mathfrak{N}} \\
        \mathbb{O}_{\mathfrak{N},3} & \mathbb{O}_{\mathfrak{N},3} & \mathbb{I}_{\mathfrak{N}} \\
    \end{pmatrix}\;,
\end{equation}
which adopts a light-cone basis along the $\mathrm{SO}(3, 3)$ coordinates, but a cartesian basis in the last $\mathfrak{N}$ directions. The matrix $M_{MN}$ can be written in terms of a vielbein $\tensor{\mathcal{V}}{_M^{\underline{M}}}$ as 
\begin{equation}
\label{eq:7d_MMN_from_VMN}
  M_{MN} = \tensor{\mathcal{V}}{_M^{\underline{M}}} \tensor{\mathcal{V}}{_N^{\underline{M}}} \;, 
\end{equation}
where $\underline{M}$ is a local $\mathrm{SO}(3) \times \mathrm{SO}(3+\mathfrak{N})$ index. The latter group is nothing but the compact part of the global symmetry \eqref{eq:GlobalSymmetry}. In particular, the first subgroup $\textup{SO}(3)$ can be identified as the $\textup{SU}(2)_R$ symmetry group of $\mathcal{N}=1$ 7D supergravity, whereas the second rotates the different vector multiplets.

The full set of $10 + 3 \mathfrak{N}$ scalars in our notation is 
\begin{equation}
\label{eq:7d_scalars}
    \Sigma = \Big\{ X, \, \varphi_1 , \, \varphi_2,\, \varphi_3,\, \chi_1,\, \chi_2,\, \chi_3,\, \chi_4,\, \chi_5,\, \chi_6,\, Y^{Ii} \Big\} \;.
\end{equation}
One possible parametrization for the coset representative can be built as follows. The nine moduli associated with the closed string sector (3 dilatons plus 6 axions) are encoded in the definition of two matrices $E$ and $B$ 
\begin{equation} \label{eq:E_B_parametrization}
    E= \begin{pmatrix}
        e^{\varphi_1 /2} & \chi_1 e^{\varphi_2 /2}  & \chi_2 e^{\varphi_3 /2} \\
        0 & e^{\varphi_2 /2} & \chi_3 e^{\varphi_3 /2} \\
        0 & 0 & e^{\varphi_3 /2} \\
    \end{pmatrix}\;, \qquad  B= \begin{pmatrix}
        0 & \chi_4 & \chi_5\\
        -\chi_4 & 0 & \chi_6 \\
        -\chi_5 & -\chi_6 & 0 \\
    \end{pmatrix}\;,
\end{equation}
and we can introduce a matrix $\tensor{A}{^I_i} \equiv \big(  Y^{Ii} \big) $ for the open string moduli. With these definitions, the explicit form for the vielbein is given by
\begin{equation}
    \tensor{\mathcal{V}}{_M^{\underline{N}}}= \begin{pmatrix}
        E^{T} & 0  & 0\\
        -C^{T} E^{T} & E^{-1} & A^{T} \\
        -A E^{T } & 0 & \mathbb{I}
    \end{pmatrix}\ ,
\end{equation}
where $C \equiv B+ \frac{1}{2} A^T A$. In this case, once we take $G=E^{-1} (E^{-1})^T $, the matrix $M_{MN}$ built from \eqref{eq:7d_MMN_from_VMN} acquires the following block structure 
\begin{equation} \label{eq:M_MN_Parametrization}
    M_{MN}= \begin{pmatrix}
        G^{-1} & -G^{-1} C & -G^{-1} A^{T} \\
        -C^{T} G^{-1} & \quad G+C^{T} G^{-1} C + A^{T} A \quad & C^{T} G^{-1} A^{T} + A^{T} \\
        -A G^{-1} & A G^{-1} C + A & \mathbb{I} + A G^{-1} A^{T} \\
    \end{pmatrix} \;.
\end{equation}
The kinetic terms for the scalar fields in the Lagrangian can be written as
\begin{equation}
    \mathcal{L}_{\textup{kinetic}} = -\frac{5}{2} X^{-2} \partial_{\mu} X\, \partial^{\mu} X + \frac{1}{16} \partial_{\mu} M_{MN}\, \partial^{\mu} M^{MN}\;.
\end{equation}
All the possible deformations of the theory are described in principle by $\theta$ and the two tensors that directly generalize those of the $\mathrm{SO}(3,3)$ case, namely
\begin{equation}
    \xi_A \;\longrightarrow \; \xi_M \;, \qquad \qquad f_{ABC} \; \longrightarrow \; f_{MNP}\; ,
\end{equation}
where $\xi_A$ and $f_{ABC}$ are the tensors defined in \eqref{eq:7d_SO3,3_embedding_f}, \eqref{eq:7d_SO3,3_embedding_xi}. 
However, since we are interested in the solutions with $\theta\neq 0$, we have to assume $\xi_M=0$ on grounds of \eqref{eq:7d_QC}. Thus, the gaugings that we consider are parametrized by the 3-form $f_{MNP}$. We would like it to describe now the possible fluxes coming from both the closed- and open-string sector of the would-be 10-dimensional supergravity before compactification. In particular, the objects we need to consider are the structure constants $\tensor{g}{_{IJ}^K}$ of the non-Abelian Yang-Mills group coming from the open strings. To this purpose, we can define the embedding tensor as
\begin{equation}
    f_{MNP} = \begin{cases}
        f_{ABC} & \text{if $[MNP]=[ABC]$,}\\
        \tensor{g}{_{IJ}^L} \delta_{LK} & \text{if $[MNP]=[IJK]$,}\\
        0 & \text{otherwise,}
    \end{cases}
\end{equation}
where we have assumed the index split \eqref{eq:7d_SO3,3+N_index}. If we also assume that the matrices $Q$, $\tilde{Q}$ entering the definition of $f_{ABC}$ have the diagonal form \eqref{eq:QQtildeMatrixForm}, the full set of parameters for the deformations of our 7D supergravity is 
\begin{equation}
    \Theta = \Big\{ \theta, \, q,\, \tilde{q},\, \tensor{g}{_{IJ}^K} \Big \}\;.
\end{equation}
Their interpretation in terms of quantities in the 10D theory is collected in Table \ref{tab:7d_Dictionary_fluxes_embedding}.

\begin{table}[h]
\centering
\setlength{\arraycolsep}{10pt} 
\renewcommand{\arraystretch}{1.5}
\(\begin{array}{|c|c|}
    \hline
        \text{Type IIA fluxes}  & \text{Embedding tensor} \\
    \hline \hline
        F_{(0)} & \sqrt{2}\; \tilde{q} \\ \hline
        H_{ijk} & \displaystyle{\frac{1}{\sqrt{2}}} \;\theta \;\varepsilon_{ijk} \\ \hline
        \Theta_{ij} & q \;\delta_{ij} \\[4pt] \hline
        \tensor{g}{_{IJ}^K} &  \tensor{f}{_{IJL}} \; \tensor{\delta}{^{LK}}\\ \hline
       
    \end{array}\)   
\caption{\emph{Dictionary between embedding tensor components and geometric fluxes of type IIA supergravity; $\Theta_{ij}$ represents the extrinsic curvature of the internal manifold} \cite{Danielsson:2017max}. }  
\label{tab:7d_Dictionary_fluxes_embedding}
\end{table}
The scalar potential of the theory, induced by the gauging, can be expressed in terms of the embedding tensor and the scalar fields as
\begin{equation} \label{eq:7d_Potential}
\begin{split}
    V=& \;\frac{1}{64} \bigg[ \frac{X^{2}}{2} f_{MNP} f_{QRS} \Big( \frac{1}{3} M^{MQ} M^{NR} M^{PS} + \Big( \frac{2}{3} \eta^{MQ} - M^{MQ}\Big) \eta^{NR} \eta^{PS} \Big) +  \\
    &+ \theta^2 X^{-8} -\frac{2 \sqrt{2}}{3} \theta X^{-3}  f_{MNP } M^{MNP} \bigg]\;,
\end{split} 
\end{equation}
which is exactly the same structure of the formula \eqref{eq:7d_SO3,3_Potential}, but with all the $\mathrm{SO}(3,3)$ indices promoted to $\mathrm{SO}(3,3+\mathfrak{N})$ indices.
The antisymmetric tensor $M^{MNP}$ is defined by using the metric \eqref{eq:Metric_SO_3_3+N} to raise the three indices of 
\begin{equation}
    M_{MNP} \equiv \varepsilon_{\underline{i} \underline{j} \underline{k} }\, \tensor{\mathring{\mathcal{V}}}{_M^{\underline{i}}}\, \tensor{\mathring{\mathcal{V}}}{_N^{\underline{j}}}\, \tensor{\mathring{\mathcal{V}}}{_P^{\underline{k}}} \;,
\end{equation}
where in the vielbein $\tensor{\mathring{\mathcal{V}}}{_M^{\underline{M}}}=\tensor{\mathcal{V}}{_M^{\underline{N}}} \tensor{R}{_{\underline{N}}^{\underline{M}}}$ the local $\mathrm{SO}(3)\times \mathrm{SO}(3+\mathfrak{N})$ index has been rotated to cartesian coordinates through a matrix
\begin{equation}
    R \equiv \frac{1}{\sqrt{2}} \begin{pmatrix}
        -\mathbb{I}_3 & \mathbb{I}_3 & \mathbb{O}_{3,\mathfrak{N}} \\
        \mathbb{I}_3 & \mathbb{I}_3 & \mathbb{O}_{3,\mathfrak{N}} \\
        \mathbb{O}_{\mathfrak{N},3} & \mathbb{O}_{\mathfrak{N},3} & \sqrt{2} \;\mathbb{I}_{\mathfrak{N}} \\
    \end{pmatrix} \;.
\end{equation}

\subsection{Fermionic shift matrices}
The fermionic spectrum of 7D $\mathcal{N}=1$ supergravity consists of: one gravitino $\psi_{\mu \alpha}$ and one dilatino $\chi_\beta$ in the gravity multiplet and $(3 + \mathfrak{N})$ gaugini $\lambda^{\alpha \underline{A}}$, one for each vector multiplet. All fermions are symplectic Majorana spinors, transforming as doublets of SU(2)$_R$ symmetry: $\alpha =1, 2$ is the corresponding fundamental index, raised and lowered with the Levi-Civita symbol in 2 dimensions. The index $\mu=0, \dots, 6$ labels the spacetime directions, whereas $\underline{A}$ is the fundamental index of $\textup{SO}(3 + \mathfrak{N})$, rotating the different vector multiplets. Let us stress that fermions only transform under the compact part of the global symmetry group \eqref{eq:GlobalSymmetry}.

Because of the gauging, the Lagrangian of our supergravity theory gets new mass terms for the fermionic fields, of the form
\begin{equation}
    e^{-1} \mathcal{L}_{\substack{\textup{fermionic}\\\textup{masses}}} =  \tensor{A}{_1^{\alpha \beta}} \bar{\psi}_{\mu \alpha} \gamma^{\mu \nu} \psi_{\nu \beta} +  \tensor{A}{_2^{\alpha \beta}} \bar{\psi}_{\mu \alpha} \gamma^{\mu } \chi_{\beta} + \tensor{{A_3}}{_{\underline{A}}_{\alpha}^{ \beta}} \bar{\psi}_{\mu \alpha} \gamma^{\mu } \lambda^{\beta \underline{A}}.
\end{equation}
The fermionic shift matrices appearing in this expression are written in \cite{Dibitetto:2015bia} for the case with $3$ vector multiplets in the $\mathrm{SL}(4)$ formalism. For the theory with $3+ \mathfrak{N}$ vector multiplets, they are expressed in terms of the embedding tensor components and coset representatives as follows 
\begin{equation}
\label{eq:7d_A1}
    \tensor{A}{_1^{\alpha \beta}}= \frac{1}{8} \Big( \theta X^{-4} \varepsilon^{\alpha \beta} + a_1  
    X \tensor{f}{_{MNP}} \tensor{\mathcal{V}}{_{\gamma \delta}^M} \tensor{\mathcal{V}}{^N ^{\alpha \gamma}} \tensor{\mathcal{V}}{^P ^{\beta \delta}}\Big)\;,
\end{equation}

\begin{equation}
\label{eq:7d_A2}
    \tensor{A}{_2^{\alpha \beta}}= \frac{1}{8} \Big( \theta X^{-4} \varepsilon^{\alpha \beta} + a_2  
    X \tensor{f}{_{MNP}} \tensor{\mathcal{V}}{_{\gamma \delta}^M} \tensor{\mathcal{V}}{^N ^{\alpha \gamma}} \tensor{\mathcal{V}}{^P ^{\beta \delta}}\Big)\;,
\end{equation}

\begin{equation}
\label{eq:7d_A3}
    \tensor{{A_3}}{_{\underline{A}}_{\alpha}^{ \beta}}= \frac{1}{8} \Big( 
    X \tensor{f}{_{MNP}} \tensor{\mathcal{V}}{^M _{\underline{A}}}\tensor{\mathcal{V}}{_{\alpha \gamma}^N} \tensor{\mathcal{V}}{^P ^{\gamma \beta}} \Big)\;.
\end{equation}
The coset representatives with  $\mathrm{SU}(2)_R$ components have been obtained from $\tensor{\mathcal{V}}{_M^{\underline{M}}}$ considering only the $\mathrm{SO}(3)$ directions of the $\mathrm{SO}(3) \times \mathrm{SO}(3+ \mathfrak{N})$ index $\underline{M}$ and rotating it with the Pauli matrices $\sigma_i$
\begin{equation}
   \tensor{\mathcal{V}}{^M ^{\alpha \beta}}= \eta^{MN} \, \tensor{\mathcal{V}}{_N^{\underline{i}}} \, \varepsilon^{\alpha \gamma} \, \tensor{(\sigma_i)}{_{\gamma}^{\beta}}\;,
\end{equation}
while $ \tensor{\mathcal{V}}{_{\alpha \beta}^M}= (\tensor{\mathcal{V}}{^M ^{\alpha \beta}})^*$ is the complex conjugate.

The tensors \eqref{eq:7d_A1}, \eqref{eq:7d_A2}, \eqref{eq:7d_A3}, together with their conjugates  $\tensor{{A_1}}{_{\alpha \beta}}=\big(  \tensor{{A_1}}{^{\alpha \beta}}\big)^*$, $\tensor{{A_2}}{_{\alpha \beta}}=\big(  \tensor{{A_2}}{^{\alpha \beta}}\big)^*$, $\tensor{{A_3}}{^{\underline{A}}^{\alpha}_{ \beta}}=\big(  \tensor{{A_3}}{_{\underline{A}}_{\alpha}^{ \beta}}\big)^*$, allow to write an alternative formula for the scalar potential
\begin{equation}
    V = -\frac{3}{10} \,\abs{A_1}^2 + \frac{4}{5} \,\abs{A_2}^2 + \frac{1}{2} \,\abs{A_3}^2 \;.
\end{equation}
Correspondingly, we can fix the numerical coefficients appearing in \eqref{eq:7d_A1}, \eqref{eq:7d_A2} as 
\begin{equation}
    a_1 = - i \, \frac{\sqrt{2}}{3}\;, \qquad a_2 = \frac{i}{\sqrt{2} \, 6}\;.
\end{equation}

For $7$D supergravity, the conditions for preserving supersymmetry in the vacuum can be expressed in terms of $A_1$ as the existence of an $\mathrm{SU}(2)_R$ spinor satisfying 
\begin{equation}
    \label{eq:7d_A1_susy}
    \tensor{A}{_1^{\alpha \beta}} q_{\beta} = \lambda \, q^{\alpha} \,, \quad \text{with}\; \abs{\lambda}=\sqrt{-\frac{5}{3} V} \;,
\end{equation}
or, in terms of $A_2$ and $A_3$, as
\begin{equation}
    \tensor{A}{_2^{\alpha \beta}} q_{\beta} =0 \; , \qquad \tensor{{A_3}}{_{\underline{A}}_{\alpha}^{ \beta}} q_{\beta} =0\;.
\end{equation}

\subsection{Coupling to \texorpdfstring{$\mathfrak{N}=3$}{N=3} extra vector multiplets and \texorpdfstring{$\mathrm{SO}(3)$}{SO(3)} truncation}
As a first case study, we consider 7D half-maximal supergravity coupled to $\mathfrak{N}=3$ extra vector multiplets. This choice is motivated by the fact that the simplest non-Abelian group needs at least three generators. In this case, we have $19$ scalar fields that parametrize a scalar manifold of the form
\begin{equation}
\label{eq:7d_N=3_scalar_manifold}
    \mathbb{R}^{+} \times \frac{\mathrm{SO}(3,6)}{\mathrm{SO}(3)\times \mathrm{SO}(6)} \; .
\end{equation}
The parametrization of the open string fluxes, in this case, has only one possible choice: 
\begin{equation} 
\label{eq:7d_gIJK}
     \tensor{g}{_{IJ}^K}\,= \tensor{\varepsilon}{_{IJL}} \; \tensor{\delta}{^{LK}}\,g_1 \quad \longrightarrow \quad f_{IJK}= \varepsilon_{IJK}\, g_1 \;.
\end{equation}
The choice of the matrices $Q$ and $\tilde{Q}$ of the form \eqref{eq:QQtildeMatrixForm}, once translated into the $\mathrm{SO}(3,3)$ representation, implies the invariance of the embedding tensor under an $\mathrm{SO}(3)$ group acting simultaneously on the $i$ and $\bar{i}$ coordinates. Due to the assumption \eqref{eq:7d_gIJK}, the $I$ sector of the embedding tensor is also $\mathrm{SO}(3)$ invariant. It is then natural to restrict our discussion to an $\mathrm{SO}(3)$ truncation of the original theory, i.\,e. assume that the only non-vanishing scalars and fluxes are the ones transforming as singlets under the $\textup{SO}(3)$ diagonal subgroup of 
$$
\mathrm{SO}(3)_i \times \mathrm{SO}(3)_{\bar{i}} \times \mathrm{SO}(3)_{I} \; \subset \; \mathrm{SO}(3,6) \; .
$$
In particular, concerning the open string sector of the scalar fields, this means that we assume they have a diagonal form
\begin{equation}
\label{eq:7d_Ydiagonal}
    Y^{Ii} = Y \delta^{Ii} \;.
\end{equation}

\section{Analysis of the vacuum structure}\label{Section:7d_vacua}
In order to investigate the possible vacuum solutions of the theory, in general we have to extremize the potential with respect to all of the scalar fields and at the same time solve the quadratic constraints for the embedding tensor to ensure consistency of the gauging. However, in the case we are considering, the assumption that $\xi_M=0$, together with the specific form of the matrices $Q$ and $\tilde{Q}$, automatically satisfies the quadratic constraints \eqref{eq:7d_QC}. Then, the system of equations to solve simply corresponds to setting to zero the derivatives of the potential with respect to the scalars \eqref{eq:7d_scalars}:
\begin{equation}
    \partial_{\Sigma} V ( \Sigma, \theta, f_{MNP} ) =0 \; .
\end{equation}

\subsection{Solutions in the origin}

Since we make use of the embedding tensor formalism, it would be very convenient to adopt the \emph{going to the origin} approach in order to simplify our problem without loss of generality. This is done for the 7D half-maximal theory with only $3$ vector multiplets (i.\,e. in absence of open string scalars and fluxes) in \cite{Dibitetto:2015bia}. However, this approach is not general here because we have defined the embedding tensor in such a way to turn on only those components that have a clear interpretation as geometric fluxes coming from the higher-dimensional theory after compactification. Hence, if we performed a generic rotation of the point of the scalar manifold and simultaneously of the embedding tensor, we would possibly turn on undesired components of the tensor.

Nevertheless, we can look for solutions in the origin of the scalar manifold, as a starting point of our analysis. In the origin, we reproduce three families of vacuum solutions that are closely related to those described in \cite{Dibitetto:2015bia}, whose features have been summarized in Table \ref{Tab:ClosedAdS7Sol}. The values of the closed string fluxes ($\theta$, $q$, $\tilde{q}$) are the same, while the new flux $g_1$ can assume any value without modifing the structure of the solution. This is the case because the components $f_{IJK}$ of the embedding tensor do not contribute to the scalar potential and its derivative in the origin.

Concerning the masses of the scalar fields, the only difference with respect to those reported in Table \ref{Tab:ClosedAdS7Sol} is that we have the new sector with $3 \mathfrak{N}$ scalars. Taking into account that, in the case of AdS vacua in 7 spacetime dimensions, the Breitenlohner-Freedman bound \cite{Breitenlohner:1982jf} determining perturbative stability of the scalar fields becomes
\begin{equation}
    \frac{m^2}{\abs{\Lambda}} \geq -\frac{3}{5} \;,
\end{equation}
the results for the three families are
\begin{itemize}
    \item family $\mathbf{1}$ has 9 scalars with normalized mass values of $\displaystyle{-\frac{8}{15}}$, then the solutions are still stable;
    \item family $\mathbf{2}$ has 9 scalars with normalized mass values of $\displaystyle{\frac{14}{5}}$;
    \item family $\mathbf{3}$ has 9 scalars with normalized mass values of $\displaystyle{-\frac{4}{5}}$, which make this family of solutions unstable because the new masses are below the BF bound.
\end{itemize}
This is in agreement with the analysis of \cite{Danielsson:2017max}, where the extra $\mathfrak{N}$ vector multiplets were introduced, but without modifying the embedding tensor structure with respect to that of the $\mathrm{SO}(3,3)$ case.


\subsection{Going out of the origin}

In order to go beyond the analysis in the origin, the first possible generalization is to keep all the closed string scalar degrees of freedom in the origin, but admit generic (out of the origin) open string scalars $Y^{Ii}$. As already specified, in order to simplify the system of equations, we assume that these scalars have the form \eqref{eq:7d_Ydiagonal}, according to the $\mathrm{SO}(3)$ truncation. 

In this setting, we find five 2-parameter families of solutions. After choosing as parameters the value of the $q$ component of the embedding tensor and the scalar field $Y$, the values of the fluxes for the first three families are reported in Table \ref{Tab:7d_Solutions_123NEW}.

\begin{table}[h]
\begingroup
\begin{center}
\setlength{\tabcolsep}{10pt} 
\renewcommand{\arraystretch}{2} 
\begin{tabular}{|c||c|c|c|c|}
        \hline
        Solution & $q$ & $\tilde{q}$ & $\theta$ & $g_1$ \\\hline\hline
        $\mathbf{1}$ & $\displaystyle{q}$ & $\displaystyle{\frac{1}{2} \; q \; (2+Y^2)}$ & $\displaystyle{\frac{q}{4}}$ & $\displaystyle{-\frac{q}{2Y}}$ \\ \hline
        $\mathbf{2}$ & $\displaystyle{q}$ & $\displaystyle{\frac{1}{14}\; q \; (-16+7 Y^2)}$ & $\displaystyle{\frac{q}{14}}$ & $\displaystyle{-\frac{q}{2Y}}$ \\ \hline
        $\mathbf{3}$ & $\displaystyle{q}$ & $\displaystyle{\frac{1}{2}\; q\; (2+Y^2)}$ & $\displaystyle{\frac{q}{2}}$ & $\displaystyle{-\frac{q}{2Y}}$ \\ \hline
\end{tabular}

\end{center}
\endgroup
\caption{\emph{The three 2-parameter families of solutions generalizing the solutions in the origin reported in Table \ref{Tab:ClosedAdS7Sol}.}}
\label{Tab:7d_Solutions_123NEW}
\end{table}

These solutions show a non-trivial interplay between the closed and open string sector, expressed in particular by the values of fluxes $\tilde{q}$ and $g_1$. Table \ref{Tab:7d_Solutions_123NEW} shows that the solutions $\mathbf{1}$, $\mathbf{2}$, $\mathbf{3}$ look quite similar to the 1-parameter solutions with the same label found in the origin (whose structure is summarized in Table \ref{Tab:ClosedAdS7Sol}). In particular, they could be \textit{formally}\textcolor{red}{\footnote{Vacua in Tables \ref{Tab:7d_Solutions_123NEW} and \ref{Tab:ClosedAdS7Sol} keep anyway being distinct: as shown momentarily, the masses of the open-string perturbations are Y-independent and are different in the two instances, also in the $Y \rightarrow 0$ limit.}} restored if we took the limit $ Y \, \to \, 0$, in order to match the expressions for the closed string fluxes, but keeping the ratio $q / Y $ constant so that $g_1$ does not diverge. 

These solutions are all Anti-de Sitter; their negative vacuum energy densities are  
\begin{equation}
    V_0 (\mathbf{1}) = -\frac{15 q^2}{1024} \quad , \quad V_0 (\mathbf{2}) = -\frac{5 q^2}{448}  \quad , \quad V_0 (\mathbf{3}) = -\frac{5 q^2}{256}  \quad,
\end{equation}
which do not depend on the value of $Y$ and coincide with the vacuum energies of the three solutions in the origin.

The stability properties of the three families of solutions can be deduced from their mass spectra. The normalized masses for all the scalar fields, with the corresponding multiplicities, are collected in Table \ref{Tab:7d_Masses_123}. We can see that the masses of the scalars in the closed string sector are the same as those for the solutions in the origin, while the masses of the scalars $Y^{Ii}$ are modified. The solutions in families $\mathbf{1}$ and $\mathbf{3}$ are stable, bacause now all the masses are above the BF bound, while in the vacua of $\mathbf{2}$ instability comes from the closed string sector.  

\begin{table}[h]
\begin{minipage}{0.28\linewidth}
\begingroup
\centering

\setlength{\arraycolsep}{10pt} 
\renewcommand{\arraystretch}{1.7} 
\(\begin{array}{|c|c|}
    \hline
        \multicolumn{2}{|c|}{\mathbf{1}}\\
    \hline \hline
        -\frac{8}{15} & 1\\ \hline
        0 & 3+ \textcolor{blue}{3}\\ \hline
        \frac{16}{15} & 5+ \textcolor{blue}{5} \\ \hline
        \frac{8}{3}  & 1+ \textcolor{blue}{1} \\
    \hline
\end{array}\) 

\endgroup
\end{minipage}
\begin{minipage}{0.42\linewidth}
\begingroup
\centering

\setlength{\arraycolsep}{10pt} 
\renewcommand{\arraystretch}{1.5} 
\(\begin{array}{|c|c|}
    \hline
        \multicolumn{2}{|c|}{\mathbf{2}}\\
    \hline \hline
        \textcolor{red}{\frac{2}{35} \big( 22 - \sqrt{1954}\big)} & 1\\ \hline
        0 & 3+ \textcolor{blue}{4}\\ \hline
        \frac{12}{5} & 5 \\ \hline
        \frac{2}{35} \big( 22 + \sqrt{1954}\big) & 1\\ \hline
        \frac{42}{5}  & \textcolor{blue}{5} \\ \hline
   
\end{array}\)   

\endgroup
\end{minipage}
\begin{minipage}{0.28\linewidth}
\begingroup
\centering

\setlength{\arraycolsep}{10pt} 
\renewcommand{\arraystretch}{1.7} 
\(\begin{array}{|c|c|}
    \hline
        \multicolumn{2}{|c|}{\mathbf{3}}\\
    \hline \hline
        0 & 8+ \textcolor{blue}{8}\\ \hline
        \frac{4}{5} & 1 \\ \hline
        \frac{12}{5}  & 1+ \textcolor{blue}{1} \\ \hline

\end{array}\) 

\endgroup
\end{minipage}

\caption{\emph{Normalized masses for the three families of solutions defined in Table \ref{Tab:7d_Solutions_123NEW}, with their multiplicities. Tachyons are highlighted in red, while the multiplicities associated to the masses of the $3 \mathfrak{N}$ open string scalars are coloured in blue.}}
\label{Tab:7d_Masses_123}
\end{table}

The residual supersymmetry in these vacua can be verified using the fermionic matrices \eqref{eq:7d_A1}, \eqref{eq:7d_A2}, \eqref{eq:7d_A3}. We find that the solutions in family $\mathbf{1}$ preserve $\mathcal{N}=1$ supersymmetry, while the others are non-supersymmetric, in analogy with their counterparts found in the origin.

In addition to the vacua discussed above, we find two 2-parameter families of solutions of the equations of motion, whose structure is summarized in Table \ref{Tab:7d_Solutions_45}. They correspond to AdS vacua, where the value of the potential is 
\begin{equation}
\begin{split}
    V_0 (\mathbf{4}) & = 5 \;q^2 \;\frac{-(56+52 Y^2 + 14 Y^4) + (4+Y^2) \sqrt{4 + 98 Y^2 + 49 Y^4}}{64 \; (16 + 7 Y^2)^2} \quad , \\
    V_0 (\mathbf{5}) & = - 5 \;q^2 \;\frac{56+52 Y^2 + 14 Y^4 + (4+Y^2) \sqrt{4 + 98 Y^2 + 49 Y^4}}{64 \; (16 + 7 Y^2)^2}\quad .
\end{split}
\end{equation}
These solutions have no analogue in the theory with $3$ vector multiplets because they come from a different branch of solutions of the equation of motion for the open string scalar $Y$. However, we will not investigate these solutions further because they all show tachyon instabilities for any values of $Y$.

\begin{table}[h]
\begingroup
\begin{center}

\setlength{\tabcolsep}{7pt} 
\renewcommand{\arraystretch}{2} 
\resizebox{\textwidth}{!}{
\begin{tabular}{|c||c|c|c|c|}
        \hline
        Solution & $q$ & $\tilde{q}$ & $\theta$ & $g_1$ \\\hline\hline
        $\mathbf{4}$ & $\displaystyle{q}$ & $\displaystyle{\frac{q}{2}\; (2+Y^2)}$ & $\displaystyle{q \;\frac{12+22 Y^2 + 7 Y^4 - (2+Y^2) \sqrt{4 + 98 Y^2 + 49 Y^4}}{2 (16 + 7 Y^2)}}$ & $\displaystyle{q \;\frac{6-  \sqrt{4 + 98 Y^2 + 49 Y^4}}{2 Y (16 + 7 Y^2)}}$ \\ [2pt]\hline
        $\mathbf{5}$ & $\displaystyle{q}$ & $\displaystyle{\frac{q}{2} \; (2+Y^2)}$ & $\displaystyle{q \;\frac{12+22 Y^2 + 7 Y^4 + (2+Y^2) \sqrt{4 + 98 Y^2 + 49 Y^4}}{2 (16 + 7 Y^2)}}$ & $\displaystyle{q \;\frac{6+  \sqrt{4 + 98 Y^2 + 49 Y^4}}{2 Y (16 + 7 Y^2)}}$ \\ [2pt]\hline
\end{tabular}
}

\end{center}
\endgroup

\caption{\emph{Two new 2-parameter families of solutions not related to those found in the origin of the scalar manifold.}}
\label{Tab:7d_Solutions_45}
\end{table}

The properties of all the solutions discussed, together with their counterparts in the theory with no extra vector multiplets, are summarized in Table \ref{Tab:7d_Properties_of_solutions}.

\begin{table}[h]
\begingroup
\begin{center}
\setlength{\tabcolsep}{7pt} 
\renewcommand{\arraystretch}{1.5} 

\begin{tabular}{|c||c|c|c|c|}
        \hline
        Solution & SUSY & \makecell{Stability for $\mathfrak{N}>0$\\ out of the origin} & \makecell{Stability\\ for $\mathfrak{N}=0$} & \makecell{Stability for $\mathfrak{N}>0$\\ in the origin} \\\hline\hline
        $\mathbf{1}$ & \textcolor{Green}{\ding{51}} & \textcolor{Green}{\ding{51}}  & \textcolor{Green}{\ding{51}}  & \textcolor{Green}{\ding{51}}  \\ \hline
        $\mathbf{2}$ & \textcolor{Red}{\ding{55}} & \textcolor{Red}{\ding{55}}  & \textcolor{Red}{\ding{55}}  & \textcolor{Red}{\ding{55}}  \\ \hline
        $\mathbf{3}$ & \textcolor{Red}{\ding{55}}  & \textcolor{Green}{\ding{51}}  & \textcolor{Green}{\ding{51}}  & \textcolor{Red}{\ding{55}}  \\ \hline
        $\mathbf{4}$ & \textcolor{Red}{\ding{55}}  & \textcolor{Red}{\ding{55}}  & - & - \\ \hline
        $\mathbf{5}$ & \textcolor{Red}{\ding{55}}  & \textcolor{Red}{\ding{55}}  & - & - \\ \hline
\end{tabular}

\end{center}
\endgroup

\caption{\emph{Summary of the residual supersymmetry and stability properties of the 7-dimensional vacuum solutions discussed.}}
\label{Tab:7d_Properties_of_solutions}
\end{table}

It might be relevant to outline that we have a family of non-supersymmetric but nevertheless stable solutions. The instabilities expected on grounds of the non-supersymmetric AdS swampland conjecture \cite{Ooguri:2016pdq, Lust:2019zwm} could in principle arise from non-perturbative decay channels, but this possibility is ruled out, for the considered set of fields, by the discussion in the following Sections. Moreover, all these solutions do not exhibit scale separation, since there are always spin-2 modes whose mass is comparable with the cosmological constant \cite{Apruzzi:2019ecr}; this means that these are not full-fledged 7D vacua.

\subsection{From flux space to field space}
As previously mentioned, the embedding tensor formalism is a powerful tool for scanning the landscape of vacua of supergravity theories with arbitrary gauge group. However, the parametrization of solutions in terms of embedding tensor components lacks an immediate physical interpretation. It is therefore interesting to investigate whether at least some of the vacuum solutions found can be rewritten as different critical points of the scalar potential of a single theory, with fixed values of the fluxes. 

We know from Subsection \ref{Subsection:7d_3_sugra} that, assuming all other parameters are fixed, varying the scalar $X$ at constant $\theta$ is equivalent to varying $\theta$ at constant $X$. In particular, for fixed fluxes corresponding to family $\mathbf{1}$, we can obtain the family $\mathbf{3}$ of solutions by moving the scalar $X$ from $X=1$ to $X=2^{-1/5}$. This suggests that the procedure can be generalized.

The relevant scalars are those that survive the $\mathrm{SO}(3)$ truncation. They parametrize a scalar manifold that, starting from \eqref{eq:7d_N=3_scalar_manifold}, is reduced to
\begin{equation}
    \mathbb{R}^+ \, \times\, \frac{\mathrm{SO}(1,2)}{\mathrm{SO}(2)}\;,
\end{equation}
corresponding to three real scalar fields. These are  $X$, $Y$ (defined in \eqref{eq:7d_Ydiagonal}), and a third scalar, coming from a linear combination of the three dilatons appearing in the parameterization \eqref{eq:E_B_parametrization}, with equal weights, i.\,e. $\varphi_1+\varphi_2+\varphi_3$. In order to drop the exponentials, we define this scalar to be
\begin{equation}
    T = e^{\frac{\varphi_1+\varphi_2+\varphi_3}{6}} \;.
\end{equation}

Starting from the solutions $\mathbf{1}$ with arbitrary $Y$ away from the origin (Table \ref{Tab:7d_Solutions_123NEW}), we fix the fluxes to the values corresponding to $Y=1$, i.\,e.
\begin{equation}
\label{eq:7d_fixed_fluxes}
    \theta=\frac{q}{4} \;, \quad \tilde{q}= \frac{3}{2} q \;, \quad g_1= -\frac{q}{2} \;.
\end{equation}
The scalar potential of this theory admits 4 critical points. Two of them, as expected, correspond to the solutions $\mathbf{1}$ and $\mathbf{3}$ of Table \ref{Tab:7d_Solutions_123NEW} with $Y=1$ (which we will denote by $\mathbf{1}'$ and $\mathbf{3}'$ from now on, to avoid confusion), while the remaining two correspond to rotations of the solutions at the origin (with closed string fluxes given in Table \ref{Tab:ClosedAdS7Sol} and $Y=0$). The values of the $\mathrm{SO}(3)$-invariant scalar fields at these critical points are reported in Table \ref{Tab:7d_Solutions_FixedFluxes}.

\begin{table}[h]
\begingroup
\begin{center}
\setlength{\tabcolsep}{10pt} 
\renewcommand{\arraystretch}{2} 
\begin{tabular}{|c||c|c|c|}
        \hline
        Solution & $X$ & $T$ & $Y$ \\\hline\hline
        $\mathbf{1}'$ & $1$ & $1$ & $1$  \\ \hline
        $\mathbf{3}'$ & $\displaystyle{2^{-1/5}}$ & $1$ & $1$  \\ \hline
        $\mathbf{1}$ & $\displaystyle{\Big(\frac{3}{2}\Big)^{1/10}}$ & $\displaystyle{\Big(\frac{2}{3}\Big)^{1/2}}$ & $0$  \\ \hline
        $\mathbf{3}$ & $\displaystyle{\frac{3^{1/10}}{2^{3/10}}}$ & $\displaystyle{\Big(\frac{2}{3}\Big)^{1/2}}$ & $0$  \\ \hline
\end{tabular}
\end{center}
\endgroup
\caption{\emph{Values of the scalar fields at the critical points of the scalar potential defined by the choice \eqref{eq:7d_fixed_fluxes}.}}
\label{Tab:7d_Solutions_FixedFluxes}
\end{table}

\begin{figure}[h]
\begin{center}
\includegraphics[width=0.6\textwidth]{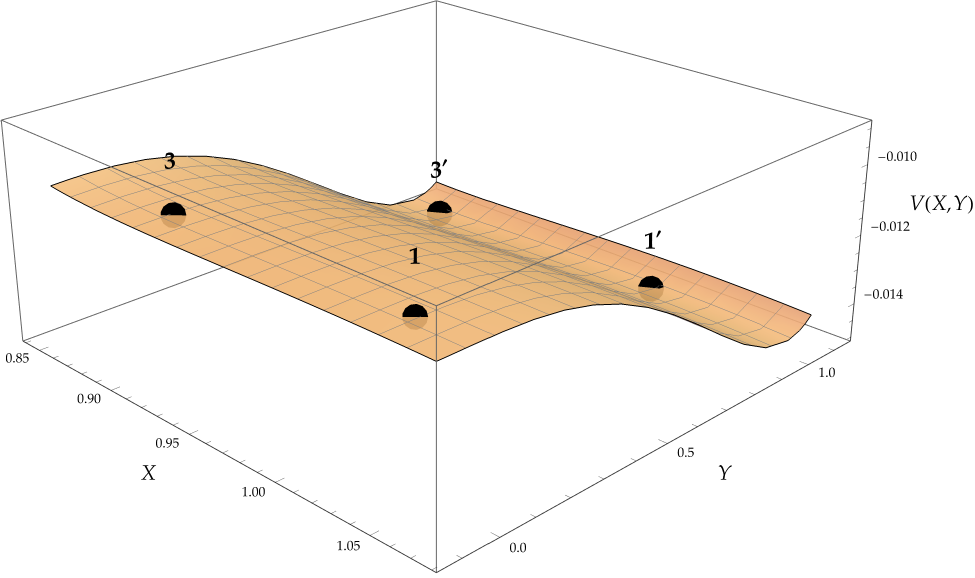}
\end{center}
\caption{\emph{Profile of the scalar potential defined by \eqref{eq:7d_fixed_fluxes}, with its critical points tabulated in \ref{Tab:7d_Solutions_FixedFluxes}. The plot is in the plane $T = \big(\frac{2}{3}\big)^{\frac{1}{2}\,(1 - Y)}$.}}
\label{Fig:7d_potential}
\end{figure}

A graph showing the profile of the scalar potential and its critical points, along the plane defined by
\begin{equation}
    T = \Big(\frac{2}{3}\Big)^{\frac{1}{2}\,(1 - Y)}\;,
\end{equation}
which contains all of the four points of Table \ref{Tab:7d_Solutions_FixedFluxes}, is shown in Figure \ref{Fig:7d_potential}. 

\section{Supergravity action and equations of motion in 10 dimensions} \label{Section:7d_10d_Reduction}
We analyze here the dimensional reduction of the type IIA supergravity action in 10 dimensions in presence of D$6$-branes. Our final aim would be in principle to compare it with the scalar potential derived from the 7D supergravity description, following a procedure similar to that followed in \cite{Balaguer:2024cyb}. 
As a further check that our 7D vacuum solutions are solutions of the full 10D supergravity, we should be also able to derive the 10D equations of motion and verify how they can be solved. 

However, as already mentioned in Section \ref{Section:WarpedCompactification}, our truncation ansatz is a warped compactification, then substituting this ansatz in the 10D action would give a non-trivial dependence on the internal coordinates (in particular the $z$ coordinate through the function $\alpha(z)$) that cannot be factored out. Then, what we can do is to start from the full ansatz with warping and extract only the first terms in the expansion of the fields around the origin of the internal manifold. In this way, we obtain by construction something independent from $z$, which we can compare with the scalar potential of 7D supergravity.

Moreover, the minimal truncation described in Section \ref{Section:WarpedCompactification} does not capture the degrees of freedom associated with the $3 + \mathfrak{N}$ vector multiplets we have included in the seven–dimensional description of Subsection \ref{Subsection:7d_3+N_sugra}. For this reason, one should not expect a complete matching between the higher- and lower-dimensional theories. The discussion presented in this paragraph should therefore be understood as outlining a procedure, namely a sequence of steps that allows one to verify the correspondence between the scalar potentials and the equations of motion. The complete matching, however, would require starting from a consistent truncation retaining the appropriate degrees of freedom.

With this caveat in mind, in the remainder of this Section we consider the truncation to minimal seven–dimensional supergravity. In order to proceed, we need some assumptions on the form of the function $\alpha(z)$ parametrizing our ansatz. As explained in \cite{Apruzzi:2019ecr}, the form of this function depends on the D6-brane configuration. In general, we can write it as
\begin{equation}
\label{eq:alpha_func}
    \alpha (z) = \alpha_0 + \alpha_1 z + \alpha_2 z^2 + \alpha_3 z^3 \;,
\end{equation}
where the coefficients $\alpha_i$ are piecewise constant, but we can always reabsorb the $\alpha_0$ in a shift of the definition of the variable $z$, so that the parameters become just three.
Moreover, since we want our metric and field configuration to be non-singular at the origin of the internal manifold (because it is the point around which we will perform our expansion, as clarified in the following Sections), we also need to assume that $\alpha_2=0$. This means having $\alpha'' (0)=0$ and, from the higher-dimensional perspective, it corresponds to the physical condition that there are no D6-branes placed in $z=0$, as discussed in \cite{Apruzzi:2019ecr}. It is then a reasonable assumption, since we want to study solutions describing dynamical branes having small fluctuations around $z=0$, parametrized by the non-vanishing scalar field $Y$.

\subsection{Bulk action}
Considering the bulk supergravity action in the democratic formulation \eqref{eq:pseudoaction}, we first look at the dimensional reduction of the Einstein-Hilbert term
\begin{equation}
\label{eq:7d_Einstein_action_10D}
    \int \dd ^{10} x \; \sqrt{-g^{(10)}} \; e^{-2 \Phi} \; \mathcal{R}^{(10)} \;.
\end{equation}
Our ansatz for the metric is the warped product \eqref{eq:WarpedMetric}, i.e. it is a metric of the form \cite{Tomasiello:2022dwe}
\begin{equation} \label{eq:7d_WarpedMetric}
    \mathrm{d}s^2_{(10)}=\, e^{2A} \, \mathrm{d}s^2_{\textup{AdS}_{7}} + \, \mathrm{d}s^2_{(3)}\;, 
 \end{equation}
where the warping function $A$ depending on the internal coordinates is, in the example of the ansatz \eqref{eq:WarpedMetric},
\begin{equation}
    A =\frac{1}{2} \log{ \Big({\sqrt{2} \pi l}\, g^2 \sqrt{-\frac{\alpha}{\alpha''}}X^{-1/2} \Big)} \;.
\end{equation}

To perform the dimensional reduction, it is convenient to rewrite the Ricci scalar $\mathcal{R}^{(10)}$ and the determinant of the metric $g^{(10)}$ respectively as
\begin{equation}
    e^{-2A} \mathcal{R}^{(\textup{AdS}_7)} +  \mathcal{R}_{ij} \, g^{ij}  \; , \qquad e^{7A} \sqrt{-g^{(\textup{AdS}_7)}} \sqrt{g^{(3)}}\; .
\end{equation}
It is worth mentioning that the term we are denoting as $\mathcal{R}_{ij}\, g^{ij}$ does \emph{not} coincide with the Ricci scalar $\mathcal{R}^{(3)}$ of the internal 3-dimensional manifold, and the difference is caused by the warping. Indeed, the components $\mathcal{R}_{ij}$ get contributions from the Riemann tensor with indices $\tensor{\mathcal{R}}{^{\mu}_{i\mu j}}$, non-vanishing due to the dependence of the spacetime part of the metric on the internal coordinates.  
Therefore, the dimensional reduction of \eqref{eq:7d_Einstein_action_10D} gives the Einstein-Hilbert term in 7 dimensions plus a contribution to the scalar potential that is
\begin{equation}
        \int \dd ^7 x \sqrt{-g^{(\textup{AdS}_7)}} \; \bigg( e^{-2\Phi}  e^{5A} \mathcal{R}^{(\textup{AdS}_7)} + \underbrace{\int \dd ^3 y \sqrt{-g^{(3)}} e^{7A} e^{-2\Phi}\mathcal{R}_{ij} \, g^{ij} }_{-V_R}\bigg)  \;,
\end{equation}
and this term shoud be evaluated at leading order in the expansion around $z=0$.

The full scalar potential arising from the bulk has other contributions coming from the fluxes over the internal manifold of the internal components of the 3-form $H_{(3)}$ and the R-R field strengths $F_{(p)}$:
\begin{equation} \label{eq:7d_VBulk}
    V_\text{Bulk}=V_R+V_{H_{3}}+\sum_p V_{F_p} \ .
\end{equation}
The contribution of the $H_{(3)}$ flux comes from
\begin{equation}
\begin{split}
    \int \dd^{10}x\, \sqrt{-g^{(10)}}\Big(& 
	-\frac{1}{12}\, e^{-2\Phi}\, |H_{(3)}|^2  \Big) \to\\
 & \to  \int \mathrm{d}^7 x\sqrt{-g^{(\textup{AdS}_7)}}\left(-\frac{1}{12}\int \dd ^3 y \sqrt{-g^{(3)}} e^{7A} e^{-2\Phi} H_{ijk}H^{ijk}\right).
  \end{split}
  \end{equation}

On the other hand, in order to compute the contributions from the fluxes of the R-R sector, we need first to discuss effects arising from the brane actions. 

\subsection{Non-Abelian brane actions}
\label{subsec:WZAction}
In order  to understand how the introduction of open string degrees of freedom contributes to the scalar potential of the supergravity theory after compactification, we consider again the non-Abelian version of Dirac-Born-Infeld (DBI) action \eqref{eq:DBI} and Wess-Zumino (WZ) action \eqref{eq:WZ} for a stack of $N_{\text{D}6}$ coincident D$6$-branes \cite{Myers:1999ps, Martucci:2005rb}. The WZ action contains topological terms involving the Ramond-Ramond fields $C_{(p)}$. While the bulk and DBI actions contribute to the 7-dimensional supergravity action through dimensional reduction, the WZ action causes a modification to the equations of motion of the fields $C_{(p)}$, or equivalently to the Bianchi identities of their duals $C_{(8-p)}$. These new Bianchi identities can be rearranged in the canonical form $\dd \widetilde{F}_{(9-p)}=0$ at the price of defining  new field strengths, modified with respect to the original ${F}_{(9-p)}$. The aim of our analysis of the WZ action is then to understand the modifications
\begin{equation}
    \Delta {F}_{(9-p)} =\widetilde{F}_{(9-p)}- {F}_{(9-p)} \;.
\end{equation}
In particular, we will be interested in the modified $F_{(0)}$ and $F_{(2)}$, since any higher order field strength would have legs in the 7-dimensional spacetime and would then break the symmetry of our vacuum solutions. This will correspond to looking at the WZ terms that involve $C_{(9)}$ and $C_{(7)}$ respectively.

The modified field strengths defined in such a way will be the ones appearing in the bulk action (then in the bulk contribution to the scalar potential \eqref{eq:7d_VBulk}), and, as a consequence, also in the bulk contribution to the equations of motion of the other fields. In particular, they will modify the graviton equations of motion and the Bianchi identity for $F_{(2)}$, as we will see in the following of this Section.

\subsubsection*{Modified $F_{(0)}$}
Modifications to the Romans' mass $F_{(0)}$ should originate from contributions to the WZ action of the form   
\begin{equation}
    \int_{10} C_{(9)} \wedge J_{(1)}^{(\textrm{D}6)} \;,
\end{equation}
or, which is equivalent up to an integration by parts, 
\begin{equation}
    \int_{10} \dd C_{(9)} \wedge \delta {\bar{F}}_{(0)} \;.
\end{equation}
The only non-vanishing term involving $C_{(9)}$ in the formal sum of the D6-brane action \eqref{eq:WZ} is
\begin{equation}
\label{WZaction_C9coupling}
     \mu_{\textrm{D}6} \int_{\textup{WV}(\textrm{D}6)}  \Tr \Big\{ \mathrm{P}  \Big[ i \lambda \iota_Y \iota_Y \big(\hat{C}_{(9)} \big)  \Big] \Big\} \;.
\end{equation}
It can be written explicitly as 
\begin{equation}
\begin{split}
   &  \mu_{\textrm{D}6} \int \dd ^7 x \Tr \Big\{ \mathrm{P}  \Big[ - \frac{1}{2} i \lambda [Y^i,Y^j] \big(\hat{C}_{(9)} \big)_{\mu_{0} \mu_{1} \mu_{2} \mu_{3} \mu_{4} \mu_{5} \mu_{6}ij}  \Big] \Big\}=\\
   =&  \mu_{\textrm{D}6} \int \dd ^7 x \Tr \Big\{ \mathrm{P}  \Big[  \frac{1}{2}  \lambda \tensor{g}{_{IJ}^K} Y^{Ii} Y^{Jj} t_K \big(\hat{C}_{(9)} \big)_{\mu_{0} \mu_{1} \mu_{2} \mu_{3} \mu_{4} \mu_{5} \mu_{6}ij}  \Big] \Big\}\;.
    \end{split}
\end{equation}
Here and in the following, we should in principle apply the pullback, but we can ignore it since the new terms generating from it would all involve spacetime derivatives of the fields $Y^i$. They contribute to the kinetic term for the $Y^i$, so they do not affect the scalar potential, nor the equations of motion if we restrict to vacuum solutions.    
The expansion of the hatted field $\hat{C}_{(9)}$ up to the relevant order in $\lambda$ is
\begin{equation}
    \big(\hat{C}_{(9)} \big)_{\mu_{0} \mu_{1} \mu_{2} \mu_{3} \mu_{4} \mu_{5} \mu_{6}ij}= \big( C_{(9)} \big)_{\mu_{0} \mu_{1} \mu_{2} \mu_{3} \mu_{4} \mu_{5} \mu_{6}ij} + \lambda Y^k \partial_k \big( C_{(9)} \big)_{\mu_{0} \mu_{1} \mu_{2} \mu_{3} \mu_{4} \mu_{5} \mu_{6}ij}
\end{equation}

The linear term in $\lambda$ vanishes once we apply the trace over the generators $t_I$, so that we are left with
\begin{equation}
\begin{split}
    & \mu_{\textrm{D}6} \int \dd ^7 x \Tr \Big\{ \frac{1}{2}  \lambda^2 \tensor{g}{_{IJ}^K} Y^{Ii} Y^{Jj} Y^{K' k} t_K t_{K'} \partial_k \big(C_{(9)} \big)_{\mu_{0} \mu_{1} \mu_{2} \mu_{3} \mu_{4} \mu_{5} \mu_{6}ij} \Big\} =\\
    = & \mu_{\textrm{D}6}  N_{\textrm{D}6}\int \dd ^7 x  \frac{1}{2}  \lambda^2 \tensor{g}{_{IJ}^K} Y^{Ii} Y^{Jj} Y^{K' k} \delta_{K K'} \partial_k \big(C_{(9)} \big)_{\mu_{0} \mu_{1} \mu_{2} \mu_{3} \mu_{4} \mu_{5} \mu_{6}ij} \; . 
    \end{split}
\end{equation}
In order to see exactly how this term contributes to the Bianchi identity for $F_{(0)}$, namely how it adds to the bulk contribution, we need to turn the 7D spacetime integral into a 10D one. This can be achieved by using Dirac delta distributions over the internal space transverse to the brane worldvolume.
\begin{equation}
\begin{split}
1= \int \dd^3 y \; \delta(\vec{y} ) = \int \dd y_1 \wedge \dd y_2 \wedge \dd y_3 \;\sqrt{\abs{g_{(3)}}} \;\frac{ \delta(y_1) \delta(y_2) \delta (y_3) }{\sqrt{\abs{g_{(3)}}}}
\end{split}
\end{equation}
Taking this into account, we can rewrite the WZ contribution to the action as the integral of a 10-form
\begin{equation}
\mu_{\textrm{D}6}  N_{\textrm{D}6}\int \dd ^{10} x \, \delta(y_1) \delta(y_2) \delta (y_3) \, \frac{1}{2}  \lambda^2 \tensor{g}{_{IJ}^K} Y^{Ii} Y^{Jj} Y^{K k} \partial_k \big(C_{(9)} \big)_{\mu_{0} \ldots \mu_{6}ij} \wedge \varepsilon_{i' j' k'} \; . 
\end{equation}
Now, we can write the full expression of the Bianchi identity for $F_{(0)}$ as 
\begin{equation}
    \frac{1}{\kappa^2_{10}} \partial_l F_{(0)} - \frac{1}{2}  \lambda^2 \mu_{\textrm{D}6}  N_{\textrm{D}6}\, \partial_l \Big(\delta(y_1) \delta(y_2) \delta (y_3) \,  \tensor{g}{_{IJ}^K} Y^{Ii} Y^{Jj} Y^{K k}\varepsilon_{ijk}\Big)=0 \;.
\end{equation}
We can rephrase this equation by defining a modified field strength
\begin{equation}
\label{eq:7d_Mod_F0}
\begin{split}    
\widetilde{F}_{(0)} =& F_{(0)} - \kappa^2_{10}  \lambda^2 \mu_{\textrm{D}6}  N_{\textrm{D}6}\, \delta(y_1) \delta(y_2) \delta (y_3) \,  \tensor{g}{_{IJ}^K} Y^{Ii} Y^{Jj} Y^{K k}\varepsilon_{ijk}  =\\
= &  - \frac{\alpha_3}{27 \pi^3}- 6 \kappa^2_{10}  \lambda^2 \mu_{\textrm{D}6}  N_{\textrm{D}6}\, \delta(y_1) \delta(y_2) \delta (y_3) \,g_1 Y^3\; 
\end{split}
\end{equation}
that satisfies a Bianchi identity in the standard form $\dd F_{(0)}=0$. In the last passage of \eqref{eq:7d_Mod_F0} we have used the ansatz \eqref{eq:ansatzF0F2} for $F_{(0)}$ together with the assumption \eqref{eq:alpha_func} on the form of the function $\alpha (z)$ and we have also restricted to the case we are considering, namely $\mathfrak{N}=3$ and a diagonal form for the fields $Y^{Ii}$.

The expression \eqref{eq:7d_Mod_F0} also agrees with the result we can get from the gauged supergravity potential: 
\begin{equation}
\label{eq:7d_Mod_F0_SUGRA}
    \widetilde{F}_{(0)}^{7D} = \tilde{q} - g_1 \, Y^3 \; .
\end{equation}

\subsubsection*{Modified $F_{(2)}$}
In order to understand the modification to the Bianchi identity for $F_{(2)}$, taking into account the duality relations \eqref{eq:duality} of type IIA democratic formulation, we have to single out contributions of the form  
\begin{equation}
\label{eq:WZ_C7_J3_current}
    \int_{10} C_{(7)} \wedge J_{(3)}^{(\textrm{D}6)} \;
\end{equation}
within the full WZ action, so that the modified Bianchi identity becomes
\begin{equation}
    \dd F_{(2)} -H_{(3)} \wedge F_{(0)}=  J_{(3)}^{(\textrm{D}6)} \;.
\end{equation}

The relevant terms in the WZ action of the D6-branes are 
\begin{equation}
\label{WZaction_C7coupling}
\begin{split}
     \mu_{\textrm{D}6} \int_{\textup{WV}(\textrm{D}6)}  \Tr \bigg\{ \mathrm{P} & \bigg[ \hat{C}_{(7)} + i \lambda \iota_Y \iota_Y \Big(\hat{C}_{(7)} \wedge \hat{B}_{(2)}\Big)  + i \lambda^2 \iota_Y \iota_Y \hat{C}_{(7)} \wedge \mathcal{F} \\
     & - \frac{\lambda^2}{2} (\iota_Y \iota_Y)^2 \Big(\hat{C}_{(7)} \wedge \hat{B}_{(2)} \wedge \hat{B}_{(2)}\Big) + \mathcal{O}(\lambda^3) \bigg] \bigg\} \quad \;.
\end{split}
\end{equation}
The last term automatically vanishes because we cannot have antisymmetric 11-forms in 10D. We can again neglect the action of the pullback, while we need the following hatted expansions
\begin{equation}
    \big(\hat{C}_{(7)} \big)_{\mu_{0} \mu_{1} \mu_{2} \mu_{3} \mu_{4} \mu_{5} \mu_{6}}= \big( C_{(7)} \big)_{\mu_{0} \ldots \mu_{6}} + \lambda Y^i \partial_i \big( C_{(7)} \big)_{\mu_{0} \ldots \mu_{6}}+ \frac{\lambda^2}{2} Y^i Y^j \partial_i \partial_j \big( C_{(7)} \big)_{\mu_{0} \ldots \mu_{6}} \;,
\end{equation}
\begin{equation}
   \big(\hat{B}_{(2)} \big)_{ij}= \lambda Y^k \partial_k B_{ij}= \lambda Y^k \frac{\big(H_{(3)} \big)_{kij}}{3} \;.
\end{equation}
In the expansion of the $\hat{B}_{(2)}$ field, the zero order in $\lambda$ is not present, according to our ansatz \eqref{eq:ansatzB2} (it is forbidden by the orientifold involution).

Then, the WZ contribution to the action can be explicitly written as 
\begin{equation}
    \begin{split}
        \mu_{\textrm{D}6} \int \dd^7 x \Tr \bigg\{  &  \big( C_{(7)} \big)_{\mu_{0} \ldots \mu_{6}} \mathbbm{1}_{N_{\textrm{D}6}} + \lambda Y^{Ii} t_I \partial_i \big( C_{(7)} \big)_{\mu_{0} \ldots \mu_{6}}+ \frac{\lambda^2}{2} Y^{Ii} Y^{Jj} t_I t_J \partial_i \partial_j \big( C_{(7)} \big)_{\mu_{0} \ldots \mu_{6}} + \\
     & + \frac{1}{2} \lambda^2 \tensor{g}{_{IJ}^K} Y^{Ii} Y^{Jj} t_K  \big( C_{(7)} \big)_{\mu_{0} \ldots \mu_{6}} Y^{K' k} t_{K'} \frac{\big(H_{(3)} \big)_{kij}}{3} \bigg\}\;.
    \end{split}
\end{equation}
The second term, linear in $\lambda$, vanishes when the trace operator is applied on it, while the third one does not contribute to the equations of motion, so that we are left with
\begin{equation}
    \mu_{\textrm{D}6} N_{\textrm{D}6}  \int \dd ^7 x \bigg\{  \big( C_{(7)} \big)_{\mu_{0} \ldots \mu_{6}} + \frac{1}{6} \lambda^2 \tensor{g}{_{IJ}^K} Y^{Ii} Y^{Jj} Y^{K' k} \delta_{K K'}  \big( C_{(7)} \big)_{\mu_{0} \ldots \mu_{6}} \big(H_{(3)} \big)_{kij} \bigg\}\;.
\end{equation}
We also need to consider here the WZ action \eqref{eq:WZOrientifold} of the orientifold planes, if present. It gives a contribution to the zero order term in $\lambda$ of the Bianchi identity 
\begin{equation}
    \mu_{\textrm{O}6} N_{\textrm{O}6}  \int \dd ^7 x   \big( C_{(7)} \big)_{\mu_{0} \ldots \mu_{6}}\;,
\end{equation}
where we know that, in the case of $\mathrm{O}6^{-}$, $\mu_{\textrm{O}6}=- 2 \, T_{\textrm{D}6}$ \cite{Gimon:1996rq, Giveon:1998sr, Bergman:2001rp}.

As in the case of $F_{(0)}$, the integrals over the 7D worldvolume can be transformed into integrals over the entire 10D manifold for the price of adding some delta functions. In this way, we can derive the new Bianchi identities for the field $F_{(2)}$ adding the contributions from the brane action to the bulk term
\begin{equation}
\begin{split}
   & \frac{1}{2 \kappa^2_{10}} \Big(\partial_k \big(F_{(2)} \big)_{ij} - \frac{1}{3}  H_{ijk} F_{(0)} \Big) + \mu_{\textrm{D}6} ( N_{\textrm{D}6} -2 N_{\textrm{O}6} ) \delta(y_1) \delta(y_2) \delta (y_3) \varepsilon_{ijk} +\\
   + & \frac{1}{6}  \lambda^2 \mu_{\textrm{D}6}  N_{\textrm{D}6}\,\delta(y_1) \delta(y_2) \delta (y_3) \,  \tensor{g}{_{IJ}^K} Y^{Ii'} Y^{Jj'} Y^{K k'} H_{i'j'k'} \; \varepsilon_{ijk} =0 \;.
    \end{split}
\end{equation}
Inserting the values of the fields and structure constants, we get the final expression for the modified Bianchi identity
\begin{multline}
\label{eq:BianchiF2Mod}
   \frac{1}{2 \kappa^2_{10}} \Big(\partial_k \big(F_{(2)} \big)_{ij} -  \frac{1}{3} H_{ijk} F_{(0)} \Big) + \\
   + \mu_{\textrm{D}6} \delta(y_1) \delta(y_2) \delta (y_3) \varepsilon_{ijk} \Big( ( N_{\textrm{D}6} -2 N_{\textrm{O}6} )
   +   \lambda^2 N_{\textrm{D}6}\, g_1 Y^3 H_{789} \Big)=0 \;.
\end{multline}

\subsubsection*{Reduction of the DBI action}
The DBI action \eqref{eq:DBI} gives also new contributions to the scalar potential of the theory and to the equations of motion of both the dilaton and the graviton fields.
In order to evaluate it, the first step is to derive the hatted expansions of all the fields, i.\,e. the metric, the $B$-field and the dilaton. Under the assumptions of the compactification ansatz in Section \ref{Section:WarpedCompactification}, they become
\begin{equation}
    \hat{g}_{\mu \nu} =\,g_{\mu \nu}+ \frac{\lambda^2}{2} Y^k Y^l \partial_k \partial_l g_{\mu \nu} =\mathbbm{1}_{1,6} \frac{g^2 L^2 l \pi ( 2 \alpha_1 + \lambda^2 3 \alpha_3 Y^2(t_I t_J \delta^{IJ}))}{2 \sqrt{3} \alpha_3 r^2 \sqrt{-\frac{\alpha_1 X}{\alpha_3}}}\,,
    \end{equation}
    \begin{multline}
    \hat{g}_{ij}=g_{ij} + \frac{\lambda^2}{2} Y^k Y^l \partial_k \partial_l g_{ij}= \sqrt{3}\sqrt{-\frac{\alpha_3}{\alpha_1}} l \pi X^{5/2} \bigg[ \delta_{ij} \Big( 2+ \frac{3 \alpha_3 t_I t_J \delta^{IJ} (-3 +8 X^5 ) Y^2 \lambda^2}{\alpha_1}\Big) +\\
    - 8 \frac{\alpha_3}{\alpha_1} t_I t_J \delta^I_i \delta^J_j (-1+3 X^5)Y^2 \lambda^2 \bigg]\,, 
    \end{multline}
    \begin{equation}
        \hat{B}_{\mu \nu} =0  \,, \qquad
    \hat{B}_{ij}=\lambda Y^k \partial_k B_{ij}= 2 \alpha_3 l \pi (-1+6 X^5 ) t_K \delta^{Kk} \varepsilon_{ijk} Y \lambda \,, 
\end{equation}
\begin{equation}
   \hat{\Phi}= \, \Phi + \frac{3 \alpha_3 t_I t_J \delta^{IJ} (-3+8 X^5) Y^2 \lambda^2}{4 \alpha_1}\,.  
\end{equation}
The expansions of the graviton and of the 2-form $B$ are used to compute $\hat{E}_{MN}=\hat{G}_{MN}+\hat{B}_{MN}$ and eventually to obtain
\begin{equation}
    \sqrt{\det \tensor{\mathbb{Q}}{^i_j}} = 1 - 2  \frac{\alpha_3}{\alpha_1} g_1 l \pi \sum_{I=1}^\mathfrak{N} {t_I}^2 (-1+3 X^5 (2+ g_1 l \pi Y)) {Y}^3 \lambda^2+ O(\lambda^3) \quad,
\end{equation}
\begin{equation}
    \sqrt{\det \tensor{\mathbb{M}}{_{MN}}} = \frac{g^7 L^7 l^{7/2} \pi^{7/2} (-\frac{\alpha_1}{\alpha_3 X})^{7/4}}{3^{7/4} r^7}-\frac{7 \alpha_1^{3/4} g^7 L^7 l^{7/2} \pi^{7/2}  \sum_{I=1}^\mathfrak{N} {t_I}^2 Y^2 \lambda^2 }{12 \, 3^{3/4} r^7 (-\alpha_3)^{3/4} X^{7/4}}+ O(\lambda^3) \quad.
\end{equation}

\subsection{Modified equations of motion}
Once we have analyzed in detail the non-Abelian brane action, we show in this Section how it contributes to the equations of motion and Bianchi identities of our theory. The starting points are the equations of motion of type IIA supergravity, summarized in Appendix \ref{Appendix:IIA_democratic}. In particular, we are dealing with equations \eqref{eq:Dilaton_eom_base} for the dilaton, \eqref{eq:Einstein_eom_base} for the graviton, while modifications to the Bianchi identity \eqref{eq:modifiedBianchiF0F2} for the field strength $F_{(2)}$ have already been described in Subsection \ref{subsec:WZAction}.

\subsubsection*{Equation for the dilaton}
Equation \eqref{eq:Dilaton_eom_base} takes into account the variation of the bulk action with respect to the dilaton. If we want to consider the contribution of the open-string sector to the dilaton equation, we simply add the variation of the DBI action, where the dilaton appears in the exponential form. The result is  
\begin{multline}
\label{eq:DilatonEOMMod}
  0=-\frac{4}{\kappa^2_{10}} \sqrt{-g^{(10)}} e^{-2 \Phi} \Big( \nabla_M \nabla^M \Phi - \nabla_M \Phi \nabla^M \Phi + \frac{1}{4} \mathcal{R} - \frac{1}{48} |H_{(3)}|^2 \Big) +\\
  +\delta(y_1) \delta(y_2) \delta (y_3)  \bigg[ T_{\textrm{D}6} \Tr \bigg( e^{-\hat{\Phi}} \sqrt{-\det(\mathbb{M}_{MN}) }\sqrt{\det(\tensor{\mathbb{Q}}{^i_j})} \; \bigg) +T_{\textrm{O}6} e^{-\Phi} \sqrt{-\det(G_{MN}) } \bigg]
\end{multline}

\subsubsection*{Einstein equations}
In order to find the modified Einstein equations, we write here the variation of the different components of the DBI action, as an intermediate step.

\begin{equation}
\begin{split}
     \delta \sqrt{-\det(\mathbb{M}_{MN}) } =& \delta \sqrt{-\det \mathrm{P} \Big[ \hat{E}_{MN} + \hat{E}_{Mi} (\mathbb{Q}^{-1} - \delta )^{ij} \hat{E}_{jN} \Big]}=\\
     = & \frac{1}{2} \sqrt{-\det(\mathbb{M}) } (\mathbb{M}^{-1})^{MN} \delta^{\mu}_M \delta^{\nu}_N \bigg[  \delta \hat{g}_{\mu \nu} + \delta \hat{b}_{\mu \nu} + \lambda \frac{\partial Y^i}{\partial x^{\mu}} (\delta \hat{g}_{i \nu} + \delta \hat{b}_{i \nu})+ \\
     & +\lambda \frac{\partial Y^i}{\partial x^{\nu}} (\delta \hat{g}_{\mu i} + \delta \hat{b}_{\mu j})+ \lambda^2 \frac{\partial Y^i}{\partial x^{\mu}} \frac{\partial Y^j}{\partial x^{\nu}} (\delta \hat{g}_{i j} + \delta \hat{b}_{i j})\bigg] + O(\lambda^3) \;,
\end{split}
\end{equation}

\begin{equation}
\begin{split}
     \delta \sqrt{-\det(\tensor{\mathbb{Q}}{^i_j}) } =& \delta \sqrt{-\det (\tensor{\delta}{^i_j}+i \lambda [Y^i,Y^k] \hat{E}_{kj} )}=\\
     = & \frac{1}{2} \sqrt{-\det(\mathbb{Q}) } \tensor{(\mathbb{Q}^{-1})}{^i_j} \Big( i \lambda [Y^j,Y^k] (\delta \hat{g}_{ki}+ \delta \hat{B}_{ki})\Big)=\\
     = &  \frac{1}{2} \sqrt{-\det(\mathbb{Q}) } \tensor{(\mathbb{Q}^{-1})}{^i_j} \Big( \lambda Y^{jJ} Y^{kK} \tensor{g}{_{JK}^I} t_I (\delta \hat{g}_{ki}+ \delta \hat{B}_{ki})\Big)\;.
\end{split}
\end{equation}

The new version of Einstein equations has to include the variation of the DBI action with respect to the metric components, but we need also two more ingredients:
\begin{itemize}
    \item Einstein equations in the form \eqref{eq:Einstein_eom_base} are obtained by replacing the 10D Ricci scalar (that would appear if we vary the Einstein-Hilbert term with respect to the metric) with its value as given by the dilaton equation of motion. Then, we need to take into account also the DBI contribution to \eqref{eq:DilatonEOMMod}.
    \item The field strengths appearing in the Ramond-Ramond sector of the bulk action should be replaced with the modified ones, as deduced from the WZ action. In our case, this only affects Romans mass $F_{(0)}$, to be replaced with $\tilde{F}_{(0)}$.
\end{itemize}
Once we define
\begin{align}
(A^{\textup{Bulk}})_{MN} = & e^{- 2 \Phi} \bigl (\mathcal{R}_{MN} + 2 \nabla_M \nabla_N \Phi -  \frac{1}{4} H_{MPQ}{H_N}^{PQ} \bigr ) - \frac{1}{2} (F_{(2)}^2)_{MN} - \frac{1}{2 \times 3!} (F_{(4)}^2)_{MN} + \notag \\  & + \frac{1}{4} G_{MN} \bigl (|\tilde{F}_{0}|^2 + \frac{1}{2!} |F_{(2)}|^2 + \frac{1}{4!}|F_{(4)}|^2 \bigr ) \; ,
\end{align}
\begin{equation}
    A^{\Phi}= \delta(y_1) \delta(y_2) \delta (y_3)  \bigg[ T_{\textrm{D}6} \Tr \bigg( e^{-\hat{\Phi}} \sqrt{-\det(\mathbb{M}_{MN}) }\sqrt{\det(\tensor{\mathbb{Q}}{^i_j})} \; \bigg) +T_{\textrm{O}6} e^{-\Phi} \sqrt{-\det(g_{MN}) } \bigg]
\end{equation}
the form of the equations for the spacetime components of the metric is
\begin{multline}
\label{eq:Graviton7DEOMMod}
    \frac{1}{2 \kappa_{10}^2} \sqrt{-g_{(10)}} (A^{\textup{Bulk}})_{\mu \nu} -\frac{1}{4} g_{\mu \nu} A^{\Phi} + \\ + \frac{1}{2} \delta(y_1) \delta(y_2) \delta (y_3)  \bigg[ T_{\textrm{D}6} \Tr \bigg( e^{-\hat{\Phi}} \sqrt{-\det(\mathbb{M}_{MN}) }\sqrt{\det(\tensor{\mathbb{Q}}{^i_j})}(\mathbb{M}^{-1})^{MN} \delta^{\rho}_M \delta^{\sigma}_N \hat{g}_{\rho \mu} \hat{g}_{\sigma \nu}\; \bigg) +\\
    +T_{\textrm{O}6} e^{-\Phi} \sqrt{-\det(g_{MN}) } g^{\rho \sigma} g_{\rho \mu} g_{\sigma \nu} \bigg]=0 \;,
\end{multline}
while for the internal components, it is
\begin{multline}
\label{eq:Graviton3DEOMMod}
    \frac{1}{2 \kappa_{10}^2} \sqrt{-g_{(10)}} (A^{\textup{Bulk}})_{ij} -\frac{1}{4} g_{ij} A^{\Phi} + \\ + \frac{1}{2} \delta(y_1) \delta(y_2) \delta (y_3)   T_{\textrm{D}6} \Tr \bigg[ e^{-\hat{\Phi}} \sqrt{-\det(\mathbb{M}_{MN}) }\sqrt{\det(\tensor{\mathbb{Q}}{^i_j})} \Big((\mathbb{M}^{-1})^{MN} \delta^{\rho}_M \delta^{\sigma}_N \lambda^2 \frac{\partial Y^k}{\partial x^{\rho}}\frac{\partial Y^l}{\partial x^{\sigma }}  +\\
    + \tensor{(\mathbb{Q}^{-1})}{^{l}_{m}}  \lambda Y^{m J} Y^{kK} \tensor{g}{_{JK}^I} t_I\Big) \hat{g}_{ki} \hat{g}_{lj} \bigg] =0 \;.
\end{multline}

\section{The Hamilton-Jacobi formulation and a positive energy theorem}
\label{Section:DW_positive_energy_theorem}

We investigate the existence of static DW solutions interpolating between AdS vacua; this class of DW solutions is usually referred to as \emph{extremal}. In a thin wall approximation, their energy density $\sigma$ is fixed by the values of the cosmological constant on each side of the wall, according to
\begin{equation} \label{eq:DW_tension_extremal}
    \sigma = \sigma_{\textup{ext}}= \frac{2}{{\kappa_{d+1}}^2} \Big( \frac{1}{L_2}- \frac{1}{L_1}\Big)\;,
\end{equation}
where $L_1$ and $L_2$ are the radii of the two AdS$_{d+1}$ extrema. The formula \eqref{eq:DW_tension_extremal} corresponds to the saturation of the Coleman-de Luccia bound \cite{Coleman:1980aw}. According to the classification of \cite{Cvetic:1992jk} (also discussed in \cite{Cvetic:1996vr}), if the tension of the wall does not fulfill this condition, we can have:
\begin{itemize}
    \item \emph{non-extremal} DWs for $\sigma > \sigma_{\textup{ext}}$, which correspond to expanding bubbles where observers from both sides of the wall are inside the bubble;
    \item \emph{ultra-extremal} DWs for $\sigma < \sigma_{\textup{ext}}$, i.\,e. solutions with an energy density is below the Coleman-de Luccia bound, which describe the evolution of a bubble of true vacuum created by the quantum tunneling process of false vacuum decay \cite{BROWN1988787}.
\end{itemize}
We will restrict our attention to \textit{flat} or \textit{planar} domain walls (DW), whose spacetime can be described as a foliation in terms of $d$-dimensional Minkowski spaces. This means that our ansatz for the metric of a DW is
\begin{equation}
\label{eq:DW_metric_general}
    \dd s^2_{(d+1)} = e^{2 A(r)} \dd s^2_{Mink_d} + e^{2B(r)} \dd r^2 \;,
\end{equation}
where $B(r)$ can always be reabsorbed by a redefinition of the coordinate $r$, but keeping it unfixed can be useful to facilitate comparison with different choices of the radial coordinate used in the literature.

The effective action describing a gravity theory in $d+1$ dimensions coupled to a set of scalar fields $\{ \phi ^I\}$ (in the Einstein frame) is
\begin{equation}
\label{eq:DW_action_general}
    \mathcal {S} = \frac{1}{ \kappa^2_{d+1}}\int{ \dd x^{d+1} \sqrt{-g} \Big(\frac{1}{2} R - \frac{1}{2} K_{IJ} \partial_{\mu} \phi^I \partial^{\mu} \phi^J - V(\phi) \Big)}\;,
\end{equation}
where $K_{IJ}$ is the kinetic metric of the scalar manifold (possibly in a non-canonical form). In the following, we will neglect the value of the gravitational constant $\kappa_{d+1}$.
If we assume a metric of the form \eqref{eq:DW_metric_general}, the Ricci scalar is
\begin{equation}
    R = - d \, e^{-2B} \big( 2 A'' + (d+1) (A')^2 - 2 A' B' \big) \;.
\end{equation}
Here and in the following, unless otherwise specified, we denote by $'$ the derivatives with respect to the radial coordinate. Replacing this expression in the action \eqref{eq:DW_action_general}  
and assuming that the scalar fields only depend on the coordinate $r$, as  $\phi^I= \phi^I (r)$, as should be the case for a generic DW solution, we can rewrite an effective 1-dimensional action. Integrating by parts to drop the second order derivatives, it becomes
\begin{equation}
    \mathcal{S}_{1\textup{D}} = \int{\dd r \,e^{dA - B} \Big[ \frac{1}{2}d (d-1) (A')^2 - \frac{1}{2} K_{IJ} {\phi^{I}}' {\phi^{J}}' - e^{2B} V(\phi)\Big]} \;.
\end{equation}
This way, the Einstein equations and the equations of motion for the scalar fields reduce to ordinary differential equations in $r$:
\begin{equation}
\label{eq:DW_eom_A_2order}
    (d-1) (A'' - A' B' ) + K_{IJ} {\phi^{I}}' {\phi^{J}}'=0 \;,
\end{equation}
\begin{equation}
\label{eq:DW_eom_phi_2order}
    K_{IJ}\,{\phi^{J}}'' + (d A' - B') K_{IJ} \, {\phi^{J}}' + \big( \frac{\partial K_{IJ}}{\partial  {\phi^{L}}} - \frac{1}{2} \frac{\partial K_{JL}}{\partial  {\phi^{I}}}\big) {\phi^{J}}' {\phi^{K}}' - e^{2B} \frac{\partial V}{\partial {\phi^I}}=0 \;.
\end{equation}
The equation of motion for $B(r)$, which is not dynamical, gives the following constraint
\begin{equation}
     \frac{1}{2} d (d-1) (A')^2 - \frac{1}{2} K_{IJ} {\phi^{I}}' {\phi^{J}}' + e^{2B} V(\phi)=0 \, ,
\end{equation}
which has been used to simplify the equation \eqref{eq:DW_eom_A_2order}.
These formulae correspond to the multi-field version of those presented in \cite{Freedman:2003ax}. 

In order to move from the Lagrangian to the Hamiltonian description of this system, we can introduce the conjugate momenta to $A$ and to the $\phi^I$
\begin{equation}
\label{eq:DW_canonical_momenta}
    \begin{cases}
        &\pi^{(A)} =\displaystyle{ \frac{\partial \mathcal{L}}{\partial A'}}= d (d-1) e^{d A-B} A' \;,\\[8pt]
        &\pi^{(\phi)}_I =\displaystyle{\frac{\partial \mathcal{L}}{\partial {\phi^I}'}}= - e^{dA-B} K_{IJ} {\phi^J}'\;.
    \end{cases}
\end{equation}
After inverting the relations \eqref{eq:DW_canonical_momenta}, the Hamiltonian functional is defined via a Legendre transform as
\begin{equation}
   \mathcal{H} [A,\,\pi^{(A)},\, \phi^I, \, \pi^{(\phi)}_I] = \frac{1}{2 \, d (d-1)} e^{B-dA}\big( \pi ^{(A)}\big) ^2 -\frac{1}{2} e^{B-dA } K^{IJ} \pi^{(\phi)}_I \pi^{(\phi)}_J + e^{B+ d A} V(\phi)\;.
\end{equation}
At this point, we can adopt the HJ formalism, which requires the introduction of the \emph{Hamilton's principal function}. Its generating functional $F$, i.\,e. the part that does not depend explicitly on $r$, must obey 
\begin{equation}
\label{eq:HJ_constraint_general}
   \mathcal{H} \Big[A,\,\frac{\partial F}{\partial A},\, \phi^I, \,\frac{\partial F}{\partial \phi^I}\Big] = \frac{1}{2\, d (d-1)} e^{B-dA}\Big( \frac{\partial F}{\partial A}\Big) ^2 -\frac{1}{2} e^{B-dA } K^{IJ}\frac{\partial F}{\partial \phi^I}\frac{\partial F}{\partial \phi^J} + e^{B+ d A} V(\phi)=0\;,
\end{equation}
which is a first order PDE for $F$. If a function $F$ satisfying \eqref{eq:HJ_constraint_general} can be found, then the dynamics of the system is reproduced solving the system of first order ordinary differential equations
\begin{equation}
    \begin{cases}
        &\pi^{(A)} =\displaystyle{ \frac{\partial F}{\partial A}} \;,\\[8pt]
        &\pi^{(\phi)}_I =\displaystyle{\frac{\partial F}{\partial {\phi^I}}}\;,
    \end{cases}
\end{equation}
In this case, the hamiltonian describing our system is such that we can use a factorized ansatz for $F$
\begin{equation}
    F[A, \, \phi^I] = e^{d A } f (\phi^I) \;.
\end{equation}
If we replace this ansatz into \eqref{eq:HJ_constraint_general}, we get the following HJ equation 
\begin{equation}
\label{eq:DW_fake_susy}
    V = -\frac{1}{2} \frac{d}{d-1} f^2 + \frac{1}{2} K^{IJ } \frac{\partial f}{\partial \phi^I}\frac{\partial f}{\partial \phi^J}\;,
\end{equation}
while the first order flows become 
\begin{equation}
\label{eq:DW_flow_1order}
    \begin{cases}
        & A' = \displaystyle{\frac{1}{d-1}} e^{B } f\;,\\[8pt]
        &{\phi^I}' = - e^{B } K^{IJ} \displaystyle{\frac{\partial f}{\partial \phi^J}}\;.
    \end{cases}
\end{equation}
The procedure of using HJ equations for a gravity theory to derive first-order RG-flow equations was initiated in \cite{deBoer:1999tgo} in the context of \emph{holographic renormalization} (also see \cite{Bianchi:2001kw, Skenderis:2002wp}).
A derivation analogous to the one of this section, but starting from generic (non-extremal) domain-walls, can be found in \cite{Danielsson:2016rmq}, where it is shown that the extremality condition is crucial to allow separability of the generating functional $F$, and eventually to derive \eqref{eq:DW_fake_susy} and \eqref{eq:DW_flow_1order}. 

A comparison between equations \eqref{eq:DW_eom_A_2order} \& \eqref{eq:DW_eom_phi_2order} and \eqref{eq:DW_flow_1order} makes it clear that the transition from a second order formalism to a first order one considerably simplifies the problem. However, the HJ treatment moves the difficulty to finding a proper generating functional. 

On the other hand, the form of \eqref{eq:DW_fake_susy} closely resembles the structure of the scalar potential of $\mathcal{N}=1$ supergravity theories in terms of the superpotential $W$ and the K\"ahler potential $K$. 
Hence, it is clear that, in the case of supergravity theories with supersymmetric AdS vacua, we can find a supersymmetric DW solution assuming that 
\begin{equation} \label{eq:Kahler}
    f_{\textup{SUSY}} \propto  e^{K/2} \, \abs{W}
\end{equation}
and inserting it into the flow equations \eqref{eq:DW_flow_1order}. 

However, the HJ equations that we have derived do not rely at all on supersymmetry. This means that we can have non-supersymmetric solutions to \eqref{eq:DW_fake_susy}, and we will show an explicit method to derive them in Section \ref{Section:DW_nonsusy_solutions}. We can denote them with $f_{\textup{FAKE}}$, which reminds that they have formally the role of a superpotential in \eqref{eq:DW_fake_susy}, but without being related to the conservation of any supersymmetry charge, i.\,e. act as a \emph{fake superpotential}. Since equation \eqref{eq:DW_fake_susy} is non-linear, we have no information a priori on the number of possible solutions. Therefore, if we start from supersymmetric vacua, we can find new branches of solutions $f_{\textup{FAKE}}$, in addition to $f_{\textup{SUSY}}$. Moreover, we can obtain solutions to \eqref{eq:DW_fake_susy} even starting from non-supersymmetric points.  

The results presented in this section are intimately correlated with the stability properties of AdS vacua of a supergravity theory. While perturbative stability is generically guaranteed by the BF bound \cite{Breitenlohner:1982jf}, stability with respect to fluctuations about a supersymmetric AdS background can be established non-perturbatively as in \cite{Gibbons:1983aq}, by
an extension of the proof by Witten \cite{Witten:1981mf} and Nester
\cite{Nester:1981bjx} of the positive energy theorem  for asymptotically flat spacetimes.
Subsequently,  Boucher \cite{Boucher:1984yx} and Townsend \cite{Townsend:1984iu} showed that these
methods apply to any AdS background solution of a
gravity plus scalar theory, irrespective of whether it is a sector of a supergravity theory. The condition that the scalar potential has to fulfill, also discussed in \cite{Skenderis:1999mm}, is exactly the \eqref{eq:DW_fake_susy}.

The name \emph{fake supergravity} for this formalism was introduced in \cite{Freedman:2003ax}, and was widely adopted thereafter. The results of the generalized Witten-Nester argument can be summarized in the following theorem, also discussed in \cite{Danielsson:2016rmq}.

\begin{Thm}[Positive energy theorem]
    Consider a theory of Einstein gravity in D dimensions coupled to a set of scalar fields, and assume that its scalar potential $V$ admits some perturbatively stable AdS critical points $\phi_0= \{\phi^I_0\}$. If there exists a globally defined function $f$ such that
\begin{align*}
     (i)\quad & \partial_I f |_{\phi=\phi_0} =0 \;,\\
     (ii)\quad&
     V = -\frac{1}{2} \frac{D-1}{D-2} f^2 + \frac{1}{2} K^{IJ } \frac{\partial f}{\partial \phi^I}\frac{\partial f}{\partial \phi^J}\;,\\
     (iii)\quad&  V \geq -\frac{1}{2} \frac{D-1}{D-2} f^2 \qquad \forall \,\phi \in \mathcal{M}_{\textup{scalar}} \;,
\end{align*}
then any other point in $\mathcal{M}_{\textup{scalar}}$ has higher energy than $\phi_0$ and hence $\phi_0$ is stable
against non-perturbative decay.
\end{Thm}
We are now going to exploit this theorem in order to investigate the non-perturbative stability of the vacua introduced in Section \ref{Section:7d_vacua}.

\section{Supersymmetric domain walls in 7 dimensions}
\label{Section:DW_susy_solution}
Considering our half-maximal supergravity theory in 7 dimensions with $3+3$ vector multiplets, after the $\mathrm{SO}(3)$ truncation there are three dynamical scalar fields left
\begin{equation}
\label{eq:DW_SO3_scalars}
    \{\phi^I\} = \{ X, T, Y\} \;.
\end{equation}
The kinetic terms for the scalars \eqref{eq:DW_SO3_scalars} are 
\begin{equation}
    \mathcal{L}_{\textup{kin}} = -\frac{5}{2 X^2} \partial_{\mu} X \,\partial^{\mu}X - \frac{3}{2 \,T^2} \partial_{\mu} T \,\partial^{\mu}T -\frac{3}{4} T^2\partial_{\mu} Y\,\partial^{\mu}Y\;,
\end{equation}
which means that we can write the kinetic matrix and its inverse as, respectively,
\begin{equation}
    K_{IJ} = \text{diag} \bigg(\frac{5}{ X^2}, \frac{3}{T^2},\frac{3}{2} T^2 \bigg)\;, \qquad K^{IJ}= \text{diag} \bigg(\frac{X^2}{5}, \frac{T^2}{3}, \frac{2}{3 T^2}\bigg)\;. 
\end{equation}
The scalar potential is given by \eqref{eq:7d_Potential}:
\begin{equation}
\begin{split}
    V= \,&\frac{1}{1024 \,X^8} \,  \big[16 \theta^2 - 48 T X^5 q \theta- 12 T^2 X^{10} q^2+ 
  12 T^4 X^{10} (q + 2 g_1 Y)^2 Y^2 + \\
  & +
  T^6 X^{10} (2 \tilde{q} - Y^2 (3 q + 4 g_1 Y))^2 + 8 T^3 X^5 (2 \tilde{q} - Y^2 (3 q + 4 g_1 Y)) \theta\big] \;.
  \end{split}
\end{equation}

As a first step, we look for an analytical function $f$ that satisfies
\begin{equation}
    V = - \frac{3}{5} f^2 + \frac{1}{2} K^{IJ } \frac{\partial f}{\partial \phi^I}\frac{\partial f}{\partial \phi^J}\;.
\end{equation}
Indeed, such a function exists and has the following form:
\begin{equation}
    f_{\textup{SUSY}} = \frac{1}{16} \Big[ \frac{2 \theta}{X^4} + 3 T  X q + 
 T^3 (-X \tilde{q} + \frac{1}{2} X Y^2 (3 q + 4 Y g_1)) 
 \Big]\;.
\end{equation}
It does not depend on the particular deformation of the theory (i.\,e. on the choice of the fluxes $\theta$, $q$, $\tilde{q}$ and $g_1$) and generalizes the superpotential adopted in \cite{Danielsson:2016rmq} (previously found in \cite{Campos:2000yu}).

It is easy to verify that $f_{\textup{SUSY}}$ is the superpotential of 7D supergravity. Indeed, with the choice of \eqref{eq:7d_fixed_fluxes} for the fluxes, we know that the critical points of the scalar potential are those showed in Table \ref{Tab:7d_Solutions_FixedFluxes}. Among them, solutions $\mathbf{1}$ and $\mathbf{1'}$ preserve $\mathcal{N}=1$ supersymmetry, and they are also the critical points of $f_{\textup{SUSY}}$
\begin{equation}
    \partial_I f_{\textup{SUSY}}|_{\mathbf{1}}=0 \;, \qquad \partial_I f_{\textup{SUSY}}|_{\mathbf{1'}}=0 \;. 
\end{equation}
A supersymmetric domain wall solution interpolating between $\mathbf{1}$ and $\mathbf{1'}$ should be found by solving the first order flow equations. 

When the fluxes are fixed as in \eqref{eq:7d_fixed_fluxes}, the scalar potential and the superpotential acquire the following expressions
\begin{equation}
\begin{split}
\label{eq_DW_potential_ofXTY}
    V= \,&\frac{q^2}{1024 \,X^8} \,  \big[1 - 12 T X^5 - 12 T^2 X^{10} + 
  12 T^4 X^{10} (1-Y)^2 Y^2 + \\
  & + 2 T^3 X^5 (3 - 3 Y^2 + 2 Y^3) +
  T^6 X^{10} (3 - 3 Y^2 + 2 Y^3)^2\big] \;.
  \end{split}
\end{equation}
\begin{equation}
    f_{\textup{SUSY}}= \frac{q [1 + 6 T X^5 - T^3 X^5 (3-3 Y^2 +2 Y^3)]}{32 X^4} \;.
\end{equation}
Since the flux $q$ is an overall factor in the formulae, we can neglect its numerical value: from now on, we will assume that it is fixed to $q=16$. After these considerations, we can explicitly write the flow equations \eqref{eq:DW_flow_1order}:
\begin{align}
     A' & = e^{ B} \frac{1 + 6 T X^5 - T^3 X^5 (3-3 Y^2 + 2 Y^3)}{10 X^4} \;, \\
     X'& = e^{ B} \frac{4-6 T X^5 + T^3 X^5 (3-3 Y^2 + 2 Y^3)}{10 X^3} \;, \\
     T' & = \frac{1}{2} e^{B} T^2 X (-2 + T^2 (3-3Y^2+2Y^3))\;,\\
     Y'& = - 2 e^{B} T X (1-Y) Y\;.
\end{align}
We can immediately observe that a particular combination of the flow equations vanishes
\begin{equation}
\label{eq:DW_A_flow}
    \Big(4 A' - \frac{X'}{X} + \frac{T'}{T} + \frac{Y'}{Y (1-Y)}\Big) =0 \;,
\end{equation}
thus signalling the presence of a first integral of the motion.
Integrating the equation \eqref{eq:DW_A_flow}, we can determine the warping function $A$ as a function of the three scalars 
\begin{equation}\label{eq:DW_warping_A}
    e^{4 A} = c_1  \frac{X(1-Y)}{T \,Y} \;,
\end{equation}
where $c_1$ is for the moment an arbitrary integration constant. 

For the remaining equations, a considerable simplification comes from assuming that our radial coordinate (usually denoted with $r$) is parametrized as an angle $\theta \in [0,\pi]$. We can now use the gauge freedom in the choice of $B(\theta)$ to set 
\begin{equation}
    Y(\theta) = \cos^2{(\theta/2)}\;, \quad \frac{\dd Y}{\dd \theta} = -\sin{(\theta/2)}\cos{(\theta/2)} \;,
\end{equation}
so that the flow equation for $Y$ is satisfied provided that
\begin{equation}
    e^{B(\theta)}=\frac{1}{T X \sin{(\theta)}} \;.  
\end{equation}
In this way, the flow equations for the scalars $X$ and $T$ become
\begin{align}
   \frac{\dd X}{\dd \theta} & = \frac{64 - 96\, T\, X^5 + \big(40-9 \cos{(\theta)}+ \cos{(3 \theta)}\big) T^3 X^5}{160\, T\, X^4 \sin{(\theta)}} \;,\\
    \frac{\dd T}{\dd \theta} & = \frac{T}{32 \sin{(\theta)}} \Big(-32 +\big(40-9 \cos{(\theta)}+ \cos{(3 \theta)}\big) T^2\Big) \;.
\end{align}
Their general solution is 
\begin{equation}
    T(\theta) = \frac{8 \cos{(\theta/2)}}{\sqrt{56 + 32 c_2 + (15 -32 c_2) \cos{(\theta)}- 6 \cos{( 2 \theta)} - \cos{(3 \theta)}}}\;,
\end{equation}

\begin{multline}
    \big(X (\theta) \big)^5= \frac{1}{2^{3}\,\sin^{2}{(\theta/2)}\cos{(\theta/2)}  }\; \cdot \\
    \cdot \;\frac{\big(58 + 3 c_3 + (45 - 32 c_2+4 c_3) \cos{(\theta)}+ (6 + c_3) \cos{( 2 \theta)} + \cos{(3 \theta)}\big) }{\sqrt{56 + 32 c_2 + (15 - 32 c_2) \cos{(\theta)}- 6 \cos{( 2 \theta)} - \cos{(3 \theta)}} } \;.
\end{multline}
The integration constants $c_2$ and $c_3$ can be fixed by imposing the boundary conditions, which correspond to the values of the scalar fields in the two supersymmetric critical points $\mathbf{1'}$ and $\mathbf{1}$
\begin{equation}
    \{X,T,Y\}_{\theta=0}=\{1,1,1\} \; , \qquad \{X,T,Y\}_{\theta=\pi}=\Big\{ \Big(\frac{3}{2}\Big)^{1/10},\Big(\frac{2}{3}\Big)^{1/2},0\Big\}\;,
\end{equation}
which are satisfied if we set $c_2=-9/16$, $c_3=-16$.

\begin{figure}[h]
\begin{center}
\includegraphics[width=0.8\textwidth]{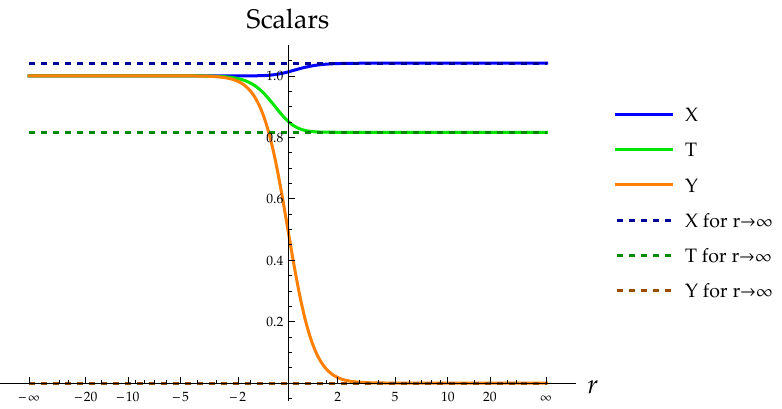}
\end{center}
\caption{\emph{Profile of the $\mathrm{SO}(3)$-invariant scalar fields along the supersymmetric domain wall interpolating between the two supersymmetric AdS vacua $\mathbf{1'}$ and $\mathbf{1}$ of the scalar potential.}}
\label{Fig:DW_susy_scalars}
\end{figure}

A more physical parametrization of our DW solution can be obtained if we choose a new radial coordinate $r$, related to the angle $\theta$ according to $\theta= 2 \arctan{(e^r)}$. In this way, the AdS vacuum $\mathbf{1'}$ sits at $r\to-\infty$, while $\mathbf{1}$ is at $r\to+\infty$. The DW profiles of the scalar fields interpolating between the two vacua are
\begin{equation}
\label{eq:DW_susy_scalars_ofr}
     \big(X(r)\big)^{10}= \frac{(2+3 \,e^{2r})^{2}}{2(2+6 \,e^{2r}+3 e^{4r})}\;, \quad \big(T(r)\big)^2= \frac{2\, (1+ e^{2r})^2}{2+6\,e^{2r}+ 3 \,e^{4r}}\;, \quad Y(r) = \frac{1}{1 + e^{2r}}\;.
\end{equation}
The warping factors $A$ and $B$ in the DW metric are:
\begin{equation}
    e^{20 \,A(r)}= e^{10 \,r} \frac{(2+3 \,e^{2r}) (2+6 \,e^{2r}+3 e^{4r})^{2}}{8 (1+ e^{2r})^{5} }\;,\quad e^{5 B(r)}= e^{-5r} \frac{(2+6 \,e^{2r}+3 e^{4r})^{3}}{2^{7} (2+3 \,e^{2r})}\;,
\end{equation}
where the integration constant $c_1$ appearing in \eqref{eq:DW_warping_A} has been fixed to $1$.
The behavior of the scalar fields along the supersymmetric domain wall, as functions of the radial coordinate $r$, is shown in Figure \ref{Fig:DW_susy_scalars}.

It is reasonable to expect that, in the two limits $r\to \pm \infty$, the domain wall metric can reproduce the metrics of the two AdS spacetimes $\dd s^2 = \dd r^2 + e^{2r/L} \dd s^2_{\textup{Mink}_d}$ with different radii $L$. Since the square radius of an Anti-de Sitter spacetime is inversely proportional to the effective cosmological constant generated by its curvature (according to $\Lambda_{\textup{eff}} = -\frac{1}{L^2} \frac{d (d-1)}{2} $), we can derive
\begin{equation}
\label{eq:DW_comparison_of _radii}
    V(\mathbf{1})=  -\frac{5}{512}\Big( \frac{3}{2}\Big)^{1/5} q^2\;, \quad V(\mathbf{1'})=-\frac{15}{1028} q^2\; \quad \rightarrow \quad \frac{L^2 (\mathbf{1})}{L^2 (\mathbf{1'})} = \Big( \frac{3}{2}\Big)^{4/5} \;.
\end{equation}
The asymptotic behaviors of the DW metric are 
\begin{equation}
\begin{split}
    \dd s^2 _{DW} & \xrightarrow{r \to -\infty} \dd s^2_{\text{AdS}\;\mathbf{1'}} = \dd r^2 + e^{r} \dd s^2_{\textup{Mink}_d}\;,\\
    \dd s^2 _{DW} & \xrightarrow{r \to +\infty} \dd s^2_{\text{AdS}\;\mathbf{1}} = \Big( \frac{3}{2}\Big)^{4/5} \dd r^2 + e^{r} \Big( \frac{3}{2}\Big)^{3/10} \dd s^2_{\textup{Mink}_d}\;.
\end{split}    
\end{equation}
If we redefine the radial coordinate in the second formula as $\tilde{r}= \big( \frac{3}{2}\big)^{2/5} r$, in such a way that the $\dd r^2$ component of the metric is canonically normalized, we get
\begin{equation}
    \dd s^2_{\text{AdS}\;\mathbf{1}} = \dd {\tilde{r}}^2 + e^{\tilde{r}  \big( \frac{2}{3}\big)^{2/5} } \Big( \frac{3}{2}\Big)^{3/10} \dd s^2_{\textup{Mink}_d}\;,
\end{equation}
where we can see explicitly that the relation between the two AdS radii in the exponential is precisely the one expected from \eqref{eq:DW_comparison_of _radii}, while the numerical factor $\displaystyle{\Big( \frac{3}{2}\Big)^{3/10}}$ can be reabsorbed in the definition of the coordinates parametrizing the d-dimensional Minkowski space.

The supersymmetric domain wall solution determined in this section was already partly discussed within a more general case in \cite{DeLuca:2018zbi}, where a class of DW solutions connecting AdS$_7$ points is interpreted as the holographic dual of RG flows between $\mathcal{N}=(1, 0)$ theories where the  flavor symmetry can be partly higgsed. However, in our case, in which we consider all of the $\mathrm{SO}(3)$ singlets as independent, we are able to give the full analytic expression of the DW flow, also thanks to a proper gauge choice for the metric.

\section{Non-supersymmetric solutions}
\label{Section:DW_nonsusy_solutions}
In this section, we will discuss a technique to perturbatively solve the PDE \eqref{eq:DW_fake_susy} and obtain a fake superpotential. Subsequently, this method will be applied to discuss other, non-supersymmetric domain wall solutions for our 7-dimensional half-maximal supergravity theory.

\subsection{Solving perturbatively the Hamilton-Jacobi equation}
As a starting point, we assume that the function $f$ satisfying \eqref{eq:DW_fake_susy} is analytical, so that it can be written as a series expansion around a critical point of the potential $\phi_0$
\begin{equation}
\label{eq:fake_f_expansion}
    f= f^{(0)} +  f^{(1)}_I \tilde{\phi}^I + \frac{1}{2!} f^{(2)}_{IJ} \,\tilde{\phi}^I\,\tilde{\phi}^J + \frac{1}{3!} f^{(3)}_{IJK} \,\tilde{\phi}^I\,\tilde{\phi}^J\,\tilde{\phi}^K+ \ldots
\end{equation}
where $\tilde{\phi}^I= (\phi^I - \phi_0^I)$ are the perturbations of the scalar fields around $\phi_0$. The scalar potential of the theory can be written in the same form
\begin{equation}
    V= V^{(0)} +  V^{(1)}_I \tilde{\phi}^I + \frac{1}{2!} V^{(2)}_{IJ} \,\tilde{\phi}^I\,\tilde{\phi}^J + \frac{1}{3!} V^{(3)}_{IJK} \,\tilde{\phi}^I\,\tilde{\phi}^J\,\tilde{\phi}^K+ \ldots
\end{equation}
but in this case the coefficients of the expansion, i.e. the derivatives of the scalar potential at $\phi=\phi_0$, are known. The other assumption we need is that $\phi_0$ is not only a critical point of the potential, but also of the fake superpotential $f$. This means that 
\begin{equation}
    V^{(1)}_I =0 \;, \quad f^{(1)}_I =0 \quad \forall I \;.
\end{equation}

Following the procedure described in \cite{Danielsson:2016rmq}, we can derive the coefficients of the expansion \eqref{eq:fake_f_expansion} order by order, taking the equation \eqref{eq:DW_fake_susy} and evaluating its derivatives at the point $\phi_0$. At the $n$-th step, we should insert in the equation all the coefficients determined in the previous $n-1$ steps, so that the only unknowns are the coefficients $f^{(n)}_{I_1 \ldots I_n}$.  
\begin{itemize}
    \item At \textbf{zero} order, after imposing that $f^{(1)}_I =0$, we have $\displaystyle{V^{(0)}= -\frac{d}{2 (d-1)}\big(f^{(0)}\big)^2}$, which is solved by
    \begin{equation}
    \label{eq:DW_HJ_perturbative_solution_zero_order}
        f^{(0)}= \sqrt{\frac{2 (d-1)}{d}V^{(0)}}\;.
    \end{equation}
    \item At order \textbf{one}, the equation 
    \begin{equation}
        V^{(1)}_I = -\frac{d}{d-1} f^{(0)} f^{(1)}_I + \frac{1}{2} (\partial_I K^{JK})_{\phi_0} f^{(1)}_J f^{(1)}_K + (K^{JK})_{\phi_0} \,f^{(2)}_{IJ} f^{(1)}_K 
    \end{equation}
    is automatically satisfied by the assumption $f^{(1)}_I =0$.
    \item At order \textbf{two}, we need to solve
    \begin{equation}
        V^{(2)}_{IJ} = -\frac{d}{d-1} f^{(0)} f^{(2)}_{IJ} + (K^{KL})_{\phi_0} \,f^{(2)}_{IK} f^{(2)}_{JL} \;.
    \end{equation}
    This system of equations can be rewritten in a matrix form. After defining $K^{1/2}$ as the unique positive-semidefinite square root of the kinetic matrix $K$ for the scalar fields, together with its inverse $K^{-1/2}$, the form of the solution is
    \begin{equation}
        K^{-1/2} f^{(2)}K^{-1/2} = \frac{d}{2(d-1)} f^{(0)} \pm \sqrt{\frac{1}{4} \Big( \frac{d}{d-1}f^{(0)} \Big)^2 + K^{-1/2} V^{(2)} K^{-1/2}}
    \end{equation}
    or, after replacing the zero-th order solution,
    \begin{equation}
\label{eq:DW_f2_coefficients}
         f^{(2)} = K \sqrt{-\frac{d}{2(d-1)} V^{(0)}}  \pm K^{1/2}\sqrt{-\frac{d}{2(d-1)} V^{(0)} + K^{-1/2} V^{(2)} K^{-1/2}} K^{1/2} \;.
    \end{equation}
    For the case of a theory with a single scalar field, this is an ordinary second order equation with two possible solutions for the derivative $f^{(2)}$. In the general case of $N$ scalars, we can interpret \eqref{eq:DW_f2_coefficients} as a system of second-degree equations for the eigenvalues of $f^{(2)}$. As a consequence, it has $2^{N}$ local branches of solutions for the set of second order derivatives $f^{(2)}_{IJ}$. It is interesting to observe that the condition allowing the $2^{N}$ solutions to be real, i.e. the positivity of the argument of the square root in the second term of \eqref{eq:DW_f2_coefficients}, is nothing but a way of rewriting the BF bound
    \begin{equation}
    \label{eq:DW_f2_coeff_BF_bound}
          \frac{K^{-1/2} V^{(2)} K^{-1/2}}{(-V^{(0)})} \geq -\frac{d}{2(d-1)}\;.
    \end{equation}
    \item After choosing one of the $2^N$ branches of solutions, for any order $k\geq 3$ we get a system of $\displaystyle{\binom{N+k-1}{k}}$ linear equations for the same number of coefficients $f^{(k)}$. This is made possible by the fact that, in any $k$-th order equation, the coefficients of order higher than $k$ (in particular, the $f^{(k+1)}$) are contracted with $f^{(1)}$, and then they have no role when we evaluate the equations at $\phi=\phi_0$, where $f^{(1)}=0$.
\end{itemize}

With this iterative procedure, we can in principle determine a solution for the function $f$ up to any  order. We will be able to obtain at most $2^{N}$ local branches of solutions starting from each critical point of the potential. However, some of these branches can show degeneracies due to non-analytic behaviors: the number of actual solutions can be then lower. 

\subsection{Fake superpotentials and more domain wall solutions}
\label{Subsec:FakeSusy_solutions}

\begin{figure}[H]
\begin{center}
\includegraphics[width=0.75\textwidth]{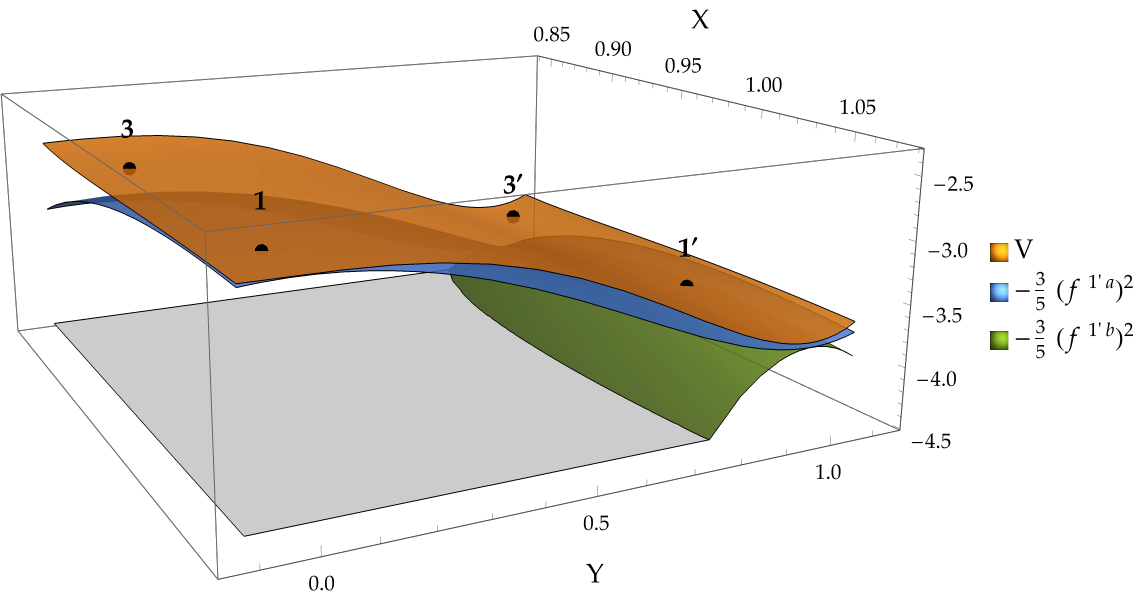}
\end{center}
\caption{\emph{Profile of the scalar potential defined by \eqref{eq:7d_fixed_fluxes}, with its critical points tabulated in \ref{Tab:7d_Solutions_FixedFluxes} and the fake superpotentials $f^{\mathbf{1'} a}$, $f^{\mathbf{1'} b}$ computed by solving \eqref{eq:DW_fake_susy} around $\mathbf{1'}$. The plot is in the plane $T = \big(\frac{2}{3}\big)^{\frac{1}{2}\,(1 - Y)}$, and we have fixed $q=16$.}}
\label{Fig:fakef_from_1'}
\end{figure}

We can apply the technique described to solve the HJ equation around the perturbatively stable critical points of the scalar potential \eqref{eq_DW_potential_ofXTY}. This means that we can look for "fake superpotentials" starting from any of the vacuum solutions $\mathbf{1'}$, $\mathbf{3'}$, and $\mathbf{1}$ of Table \ref{Tab:7d_Solutions_FixedFluxes}; we are not allowed to do the same for the solution $\mathbf{3}$ because it is unstable in the direction parametrized by the scalar field $Y$. We collect here the main results of the computation, while more details about the local branches of solutions around each point, both the analytic and non-analytic ones, are provided in Appendix \ref{Appendix:HJ_branches}.
\begin{itemize}
\item Out of the 8 branches of solutions starting from the supersymmetric vacuum $\mathbf{1'}$, 6 are not analytical. The two analytical solutions, as well as the ones starting from the other critical points that will be discussed in the following, have been determined up to order 20, in order to obtain a good numerical precision. A first branch, which we denote with $f^{\mathbf{1'} a}$, shows another critical point at $\mathbf{1}$: it reproduces the exact superpotential determined analytically in Section \ref{Section:DW_susy_solution}. A second one, called $f^{\mathbf{1'} b}$, has a global paraboloid behavior, with a minimum at $\mathbf{1'}$: it provides a lower bound for the scalar potential, satisfying the requirements of the positive energy theorem in Section \ref{Section:DW_positive_energy_theorem}. A plot of the solutions can be found in Figure \ref{Fig:fakef_from_1'}.
\end{itemize}

\begin{figure}[H]
\begin{center}
\includegraphics[width=0.75\textwidth]{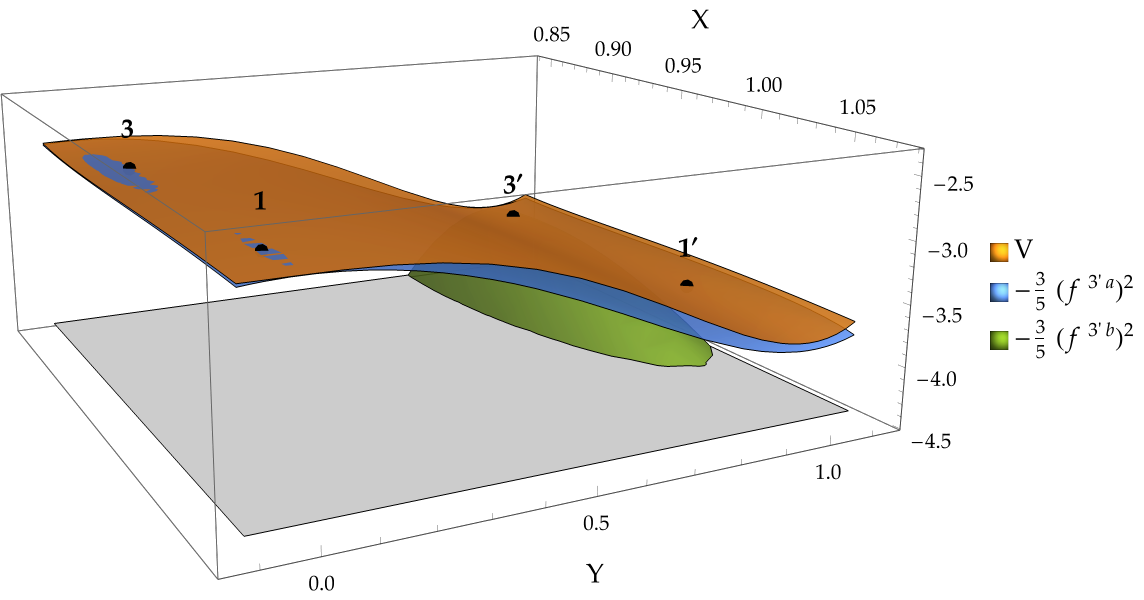}
\end{center}
\caption{\emph{Profile of the scalar potential defined by \eqref{eq:7d_fixed_fluxes}, with its critical points tabulated in \ref{Tab:7d_Solutions_FixedFluxes} and the fake superpotentials $f^{\mathbf{3'} a}$, $f^{\mathbf{3'} b}$ computed by solving \eqref{eq:DW_fake_susy} around $\mathbf{3'}$. The plot is in the plane $T = \big(\frac{2}{3}\big)^{\frac{1}{2}\,(1 - Y)}$, and we have fixed $q=16$.}}
\label{Fig:fakef_from_3'}
\end{figure}

\begin{itemize}
\item Taking as a starting point the non-supersymmetric vacuum $\mathbf{3'}$, again we find 6 branches with non-analytic behaviors. The relevant branches are $f^{\mathbf{3'} a}$, which has critical points also in $\mathbf{1'}$ and $\mathbf{1}$, then is a good candidate for solving the flow equations and finding new domain wall solutions, and $f^{\mathbf{3'} b}$, which is a paraboloid. The situation is shown in Figure \ref{Fig:fakef_from_3'}.
\end{itemize}

\begin{figure}[H]
\begin{center}
\includegraphics[width=0.75\textwidth]{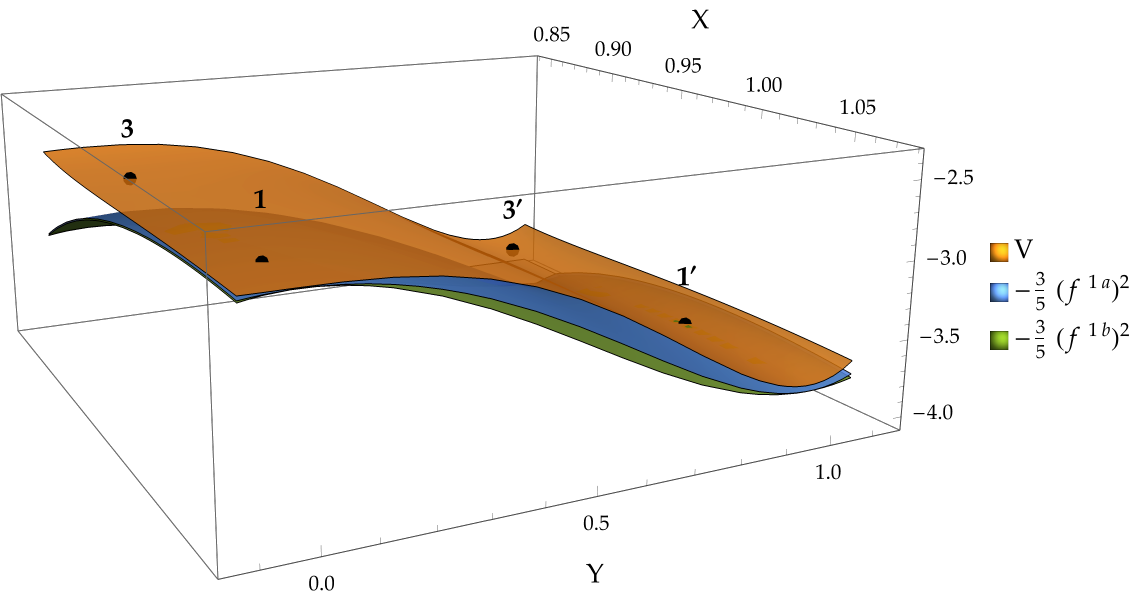}
\end{center}
\caption{\emph{Profile of the scalar potential defined by \eqref{eq:7d_fixed_fluxes}, with its critical points tabulated in \ref{Tab:7d_Solutions_FixedFluxes} and the fake superpotentials $f^{\mathbf{1} a}$, $f^{\mathbf{1} b}$ computed by solving \eqref{eq:DW_fake_susy} around $\mathbf{1}$. The plot is in the plane $T = \big(\frac{2}{3}\big)^{\frac{1}{2}\,(1 - Y)}$, and we have fixed $q=16$.}}
\label{Fig:fakef_from_1}
\end{figure}

\begin{itemize}
\item Starting from the supersymmetric vacuum $\mathbf{1}$, we find 6 branches of non-analyticity; the remaining two solutions, shown in Figure \ref{Fig:fakef_from_1} are very similar between each other and to the supersymmetric solution of the flow equations.
\end{itemize}

\begin{figure}[H]
\begin{center}
\includegraphics[width=0.8\textwidth]{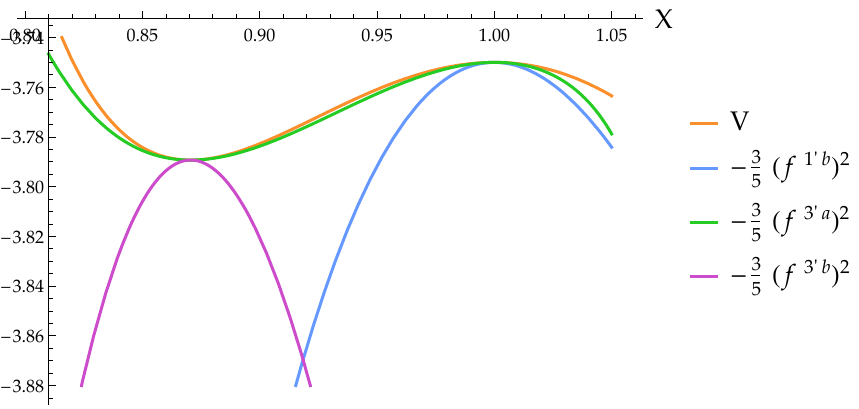}
\end{center}
\caption{\emph{Profile of the scalar potential defined by \eqref{eq:7d_fixed_fluxes}, together the function defining with the non-supersymmetric domain wall interpolating between the AdS vacua $\mathbf{3'}$ (non-supersymmetric, on the left side) and $\mathbf{1'}$ (supersymmetric, on the right) in green, and two globally bounding function ensuring non-perturbative stability of the two vacua (in blue and purple). The plot is along the line $\big(T=1,\, Y=1\big)$, and we have fixed $q=16$.}}
\label{Fig:DW_Plot2D_branches}
\end{figure}

\begin{figure}[H]
\begin{center}
\includegraphics[width=0.7\textwidth]{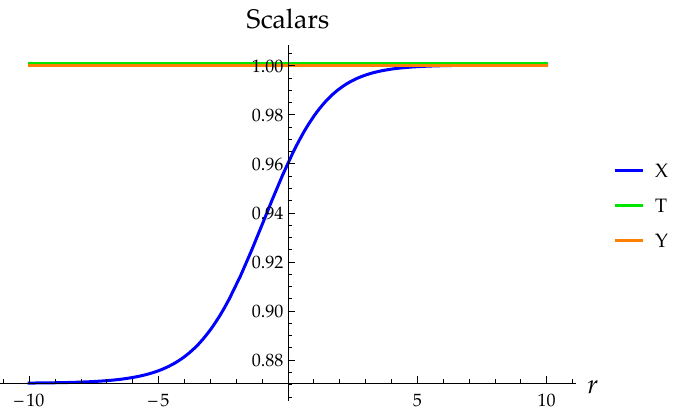}
\end{center}
\caption{\emph{Profile of the $\mathrm{SO}(3)$-invariant scalar fields along the non-supersymmetric domain wall interpolating between the AdS vacua $\mathbf{3'}$ (in the limit $r\to -\infty$) and $\mathbf{1'}$ (in the limit $r\to +\infty$). For this domain wall,  only the scalar $X$ gets a non-constant profile.}}
\label{Fig:DW_nonsusy_scalars}
\end{figure}

A 2-dimensional plot of some of the perturbative solutions determined, as in Figure \ref{Fig:DW_Plot2D_branches}, shows a projection along the $Y=1$ direction. The two global bounding functions  $f^{\mathbf{1'} b}$ and $f^{\mathbf{3'} b}$ ensure non-perturbative stability of the two vacua, while $f^{\mathbf{3'} a}$ corresponds to a static interpolating domain wall. We can also numerically solve the flow equations \eqref{eq:DW_flow_1order} with $f= f^{\mathbf{3'} a}$ to find the profile of the scalar fields along this non-supersymmetric domain wall. In this case, only the scalar $X$ is dynamical, while $T$ and $Y$ are constant along the flow. The result of integration is shown in Figure \ref{Fig:DW_nonsusy_scalars}.

The non-supersymmetric domain wall solution found from the branch $f^{\mathbf{3'} a}$ in our case has no counterpart at $Y=0$, i.\,e. we cannot find a domain wall interpolating between the AdS vacua $\mathbf{1}$ and $\mathbf{3}$, because the solution $\mathbf{3}$ does not satisfy the BF bound: it is unstable in the direction corresponding to the field $Y$. However, a domain wall solution can be found for the half-maximal supergravity theory with only three vector multiplets, whose dynamical scalar fields after the $\mathrm{SO}(3)$ truncation are only $X$ and $T$. This domain wall solution has been studied in \cite{Danielsson:2016rmq}, and previously in \cite{Campos:2000yu}. The results obtained by solving perturbatively the HJ equation are completely analogous to the ones represented in Figure \ref{Fig:DW_Plot2D_branches}: there exist two globally bounding functions (one starting from $\mathbf{1}$, the other from $\mathbf{3}$) ensuring non-perturbative stability of the two vacua along the directions of $X$ and $T$, and another branch starting from $\mathbf{3}$ that defines the non-supersymmetric domain wall.

\begin{figure}[H]
\begin{center}
\includegraphics[width=0.8\textwidth]{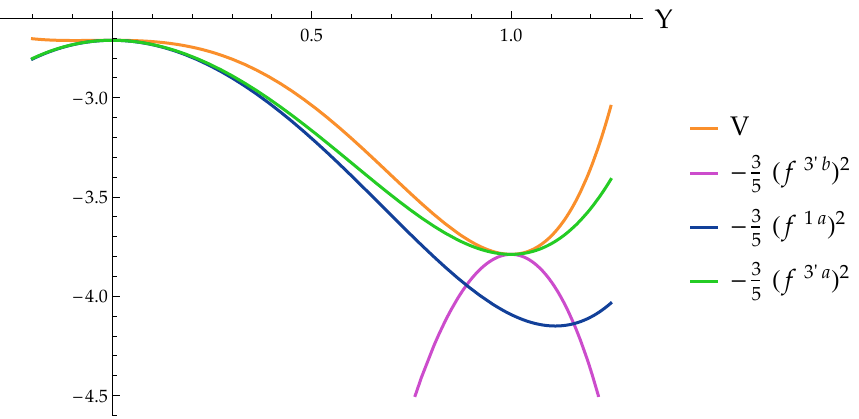}
\end{center}
\caption{\emph{Profile of the scalar potential defined by \eqref{eq:7d_fixed_fluxes}, together the function defining with the non-supersymmetric domain wall interpolating between the AdS vacua $\mathbf{3'}$ (non-supersymmetric, on the right side) and $\mathbf{1}$ (supersymmetric, on the left) in green, and two globally bounding function ensuring non-perturbative stability of the two vacua (in dark blue and purple). The plot is along the line of the $(X,T,Y)$ plane containing the two vacua $\mathbf{3'}$ and $\mathbf{1}$, which corresponds to $\big(T = \big(\frac{2}{3}\big)^{\frac{1}{2}\,(1 - Y)},\, X=\big(\frac{3}{2}\big)^{\frac{1}{10}} 6^{-\frac{Y}{10}}\big)$, and we have fixed $q=16$.}}
\label{Fig:DW_Plot2D_branches_Diagonal}
\end{figure}

Finally, we can study a third domain wall solution, interpolating between the non-supersymmetric vacuum $\mathbf{3'}$ and the supersymmetric $\mathbf{1}$. The profiles of the relevant branches of solutions of the HJ equation are shown in Figure \ref{Fig:DW_Plot2D_branches_Diagonal} along a proper line of the $(X,T,Y)$ space. 

In this case all of the three scalar fields change their value along the domain wall flow. For this reason, in order to correctly integrate the first order equations \eqref{eq:DW_flow_1order}, it is crucial to adopt a proper choice of the initial conditions. 

\begin{figure}[H]
\begin{center}
\includegraphics[width=0.8\textwidth]{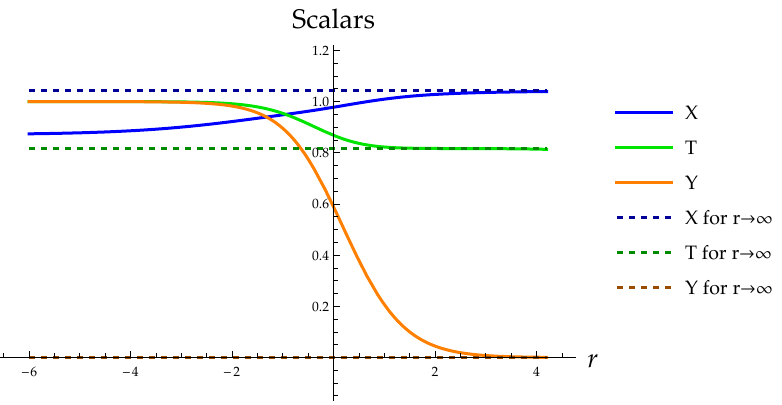}
\end{center}
\caption{\emph{Profile of the $\mathrm{SO}(3)$-invariant scalar fields along the non-supersymmetric domain wall interpolating between the AdS vacua $\mathbf{3'}$ (in the limit $r\to -\infty$) and $\mathbf{1}$ (in the limit $r\to +\infty$). }}
\label{Fig:DW_nonsusy_2_scalars}
\end{figure}

This can be done by performing a \emph{linearized boundary analysis}, similar to the one discussed in \cite{Bomans:2021ara} and used to find bubble geometries. The goal is to determine a solution that reproduces asymptotically the AdS vacuum and satisfies linearized equations of motion around it. This asymptotic solution will be used as initial conditions for the numerical integration.
The starting point is to assume the following form for the metric and the scalar fields    
\begin{align}
  e^{A} & = e^{r/L} \big( 1 + \varepsilon\, a(r) + O(\varepsilon^2)  \big)    \;,\\
  \phi^I & = \phi^I_0 + \varepsilon\, \varphi^I(r) + O(\varepsilon^2)      \;,
\end{align}
where $\varepsilon$ is a small parameter controlling the expansion. 
We also assume that we have fixed the gauge $B(r)=0$, and that the function $f$ and the inverse kinetic metric $K^{IJ}$ can be expanded perturbatively around $\phi_0$. In this way, using the identity $A'= e^{-A} (e^A)'$, the system of equations \eqref{eq:DW_flow_1order} becomes
\begin{equation}
\begin{cases}
    & \displaystyle{\frac{1 + \varepsilon a}{L} + \varepsilon  a' =\frac{1}{d-1}( 1 + \varepsilon a ) \,f(\phi_0) + O(\varepsilon^2) \;,}\\[6pt]
    & \varepsilon \,{\varphi^I }' = - \Big(K^{IJ} (\phi_0 ) + \partial_L (K^{IJ} (\phi_0 ) \varepsilon \varphi^L \Big) \Big(f^{(2)}_{JL} (\phi_0) \varepsilon \varphi^L\Big) + O(\varepsilon^2)\;.
\end{cases}
\end{equation}
At zero order in $\varepsilon$, this equation corresponds to the constraint
\begin{equation}
\label{eq:linearized_boundary_zero_order}
    f(\phi_0) = \frac{d-1}{L}\;.
\end{equation}
Comparing this equation with \eqref{eq:DW_HJ_perturbative_solution_zero_order}, we find the usual relation between the cosmological constant and the AdS radius. At first order in $\varepsilon$, since the equations must hold for any value of the perturbative parameter, we find
\begin{equation}
\begin{cases}
    & \displaystyle{\frac{a}{L} +  a' -\frac{a}{d-1}f(\phi_0) =0 \;,}\\[6pt]
    & {\varphi^I }' + K^{IJ} (\phi_0 ) \, f^{(2)}_{JL} (\phi_0) \, \varphi^L =0\;.
\end{cases}
\end{equation}
The first equation, together with the condition \eqref{eq:linearized_boundary_zero_order}, always implies that $a(r)= \text{constant}$ at first order in $\varepsilon$, while for the second equation we need to replace a solution of \eqref{eq:DW_f2_coefficients}.
In our case, however, due to the diagonal form of the kinetic matrix, the differential equations for $\varphi^I$ are decoupled, then this analysis is not conclusive, because the initial conditions still depend on three arbitrary integration constants, one for each scalar field. The result of numerical integration is shown in Figure \ref{Fig:DW_nonsusy_2_scalars}.

\begin{figure}[h]
\begin{center}
\includegraphics[width=0.8\textwidth]{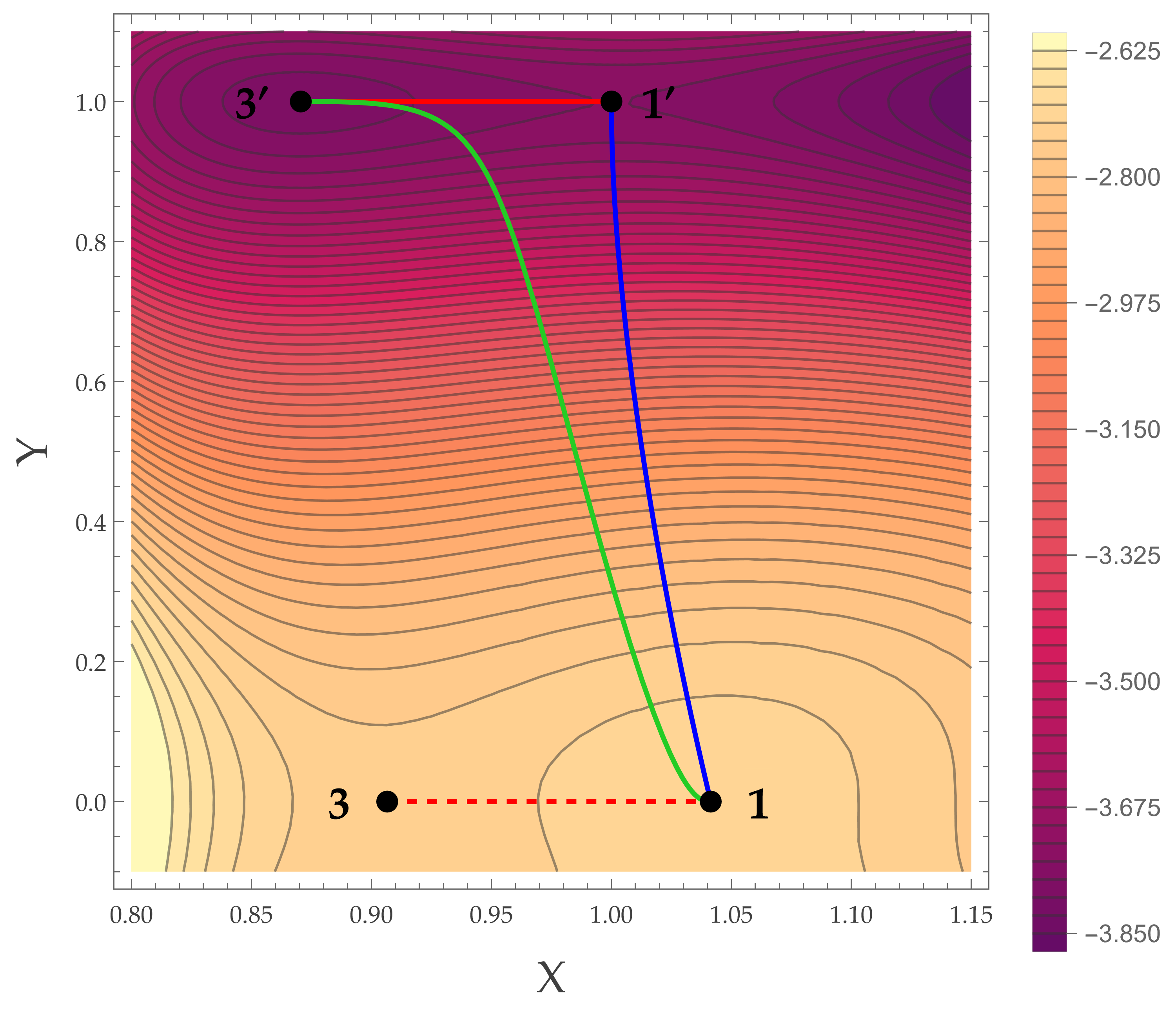}
\end{center}
\caption{\emph{Plot of the level curves of the scalar potential defined by \eqref{eq:7d_fixed_fluxes}, with its critical points tabulated in Table \ref{Tab:7d_Solutions_FixedFluxes} and the flows of the scalar fields along the domain walls. The blue curve corresponds to the only supersymmetric domain wall determined (the flow of the scalar field is represented in Figure \ref{Fig:DW_susy_scalars}). The red curve represents the non-supersymmetric DW flow from $\mathbf{3'}$ to $\mathbf{1'}$ (see Figure \ref{Fig:DW_nonsusy_scalars}), while the red dashed curve stands for the non-supersymmetric domain wall between $\mathbf{3}$ and $\mathbf{1}$, which exists only for the theory without the scalar $Y$. The flow of the scalars along non-supersymmetric DW from $\mathbf{3'}$ to $\mathbf{1}$ (see Figure \ref{Fig:DW_nonsusy_2_scalars}) corresponds to the green curve. }}
\label{Fig:DW_contour_plot}
\end{figure}

The flows of the scalar fields along the three domain wall solutions determined are summarized in the plot in Figure \ref{Fig:DW_contour_plot}. 

\section{Concluding remarks} \label{Section:conclusion}

In this work, we have studied half-maximal supergravity theory in 7-dimensions and its relation with warped compactifications of massive type IIA supergravity. For the supergravity models investigated, the outcome has been the selection of new vacuum solutions. Their features are expected to come from the presence of dynamical open strings in the corresponding 10-dimensional theory. This is strongly suggested by the analogy with the 4-dimensional vacua studied in \cite{Balaguer:2024cyb}. The lack of a consistent truncation Ansatz from 10D to 7D with an arbitrary number of vector multiplets hinders a final proof of this hypothesis; however, supergravity turns out to be an effective tool to explore the 7D landscape when dimensional reduction is not available.

A natural continuation of this work would be the search for consistent truncations to $\mathrm{AdS}_7$ of type IIA supergravity theory in 10 dimensions. Indeed, in presence of a truncation involving $3$ vector multiplets (or at least their $\mathrm{SO}(3)$-invariant sector), the application of the procedure discussed in Section \ref{Section:7d_10d_Reduction} would allow complete matching between the 10D and 7D descriptions, similarly to \cite{Balaguer:2023jei, Balaguer:2024cyb, Arboleya:2025lwu}. In that case, we could confirm that the additional degrees of freedom introduced in the supergravity theory describe an open string sector and that the vacua can be lifted to 10 dimensions. The search for new consistent truncations could be carried out primarily with the aid of exceptional field theory tools.

\begin{figure}[h]
\begin{center} 
\begin{tikzpicture}[>=latex]

\draw[-{Stealth[length=3mm]}] (-4,-3)--(-4,4);
\node (A) at (-4,-2) {$\bullet$};
\node (B) at (-5,-2) {$Y=0$};
\node (C) at (-4,3) {$\bullet$};
\node(D) at (-5,3) {$Y=1$};
\node(E) at (-5.5,2.5) {\scriptsize{(open string effect)}};
\node(F) at (-2,-2) {$\circ$};
\node(G) at (3,-2) {$\bullet$};

\draw[-{Stealth[length=3mm]},dashed] (-1.9,-2)--(0.5,-2);
\draw[dashed] (0.5,-2)--(3,-2);

\node(H) at (-2,3) {$\bullet$};
\node(I) at (3,3) {$\bullet$};

\draw[-{Stealth[length=3mm]}] (-2,3)--(0.5,3);
\draw[] (0.5,3)--(3,3);
\draw[-{Stealth[length=3mm]}] (3,3)--(3,0.5);
\draw[] (3,0.5)--(3,-2);
\draw[-{Stealth[length=3mm]}] (-2,3)--(0.5,0.5);
\draw[] (0.5,0.5)--(3,-2);

\node[draw=white,fill=white, rectangle,align=center](L) at (-2,3.5) {\scriptsize{Non-SUSY
    }};
\node[draw=none,fill=none, rectangle,align=center](L) at (-2,-2.7) {\scriptsize{Non-SUSY}\\[-1pt]\scriptsize{unstable
    }};
\node[draw=white,fill=white, rectangle,align=center](L) at (0.5,-2.8) {\scriptsize{DW non-SUSY}\\[-1pt]\scriptsize{(only for the }\\[-5pt] \scriptsize{theory without Y)
    }};
\node[draw=white,fill=white, rectangle,align=center](L) at (3,-2.5) {\scriptsize{SUSY
    }};
\node[draw=none,fill=none,rectangle,align=center](L) at (-0.5,0.5) {\scriptsize{DW}\\[-2pt] \scriptsize{non-SUSY
    }}; 
\node[draw=none,fill=none, rectangle,align=center](L) at (4,0.5) {\scriptsize{DW}\\[-2pt] \scriptsize{SUSY
    }}; 
\node[draw=none,fill=none, rectangle,align=center](L) at (0.5,3.5) {\scriptsize{DW}\\[-2pt] \scriptsize{non-SUSY
    }};
\node[draw=none,fill=none, rectangle,align=center](L) at (3,3.5) {\scriptsize{SUSY
    }};

\end{tikzpicture} 

\end{center}
\caption{\emph{Picture of the AdS$_7$ vacua of half-maximal supergravity and the domain wall solutions among them.}}
\label{Fig:Vacua_domain_walls}
\end{figure}
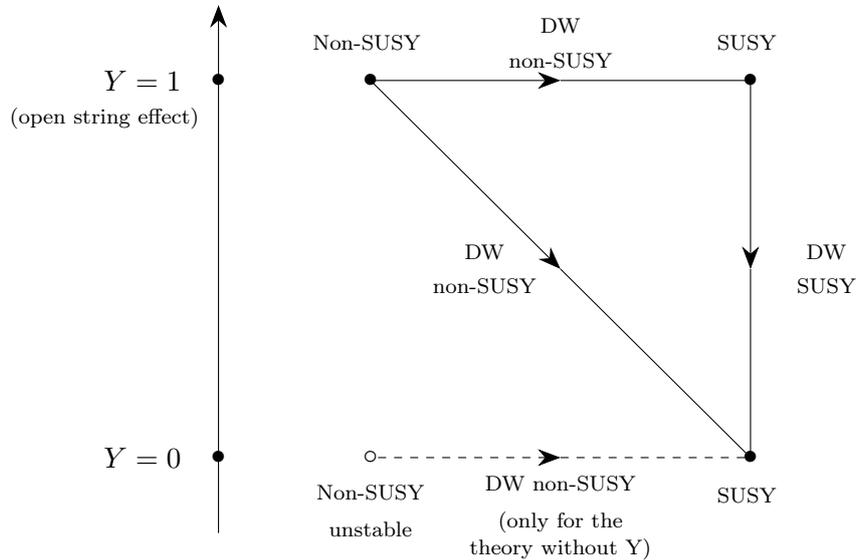

The analysis of the 7D supergravity theory has been completed in the second part of the paper, where we have addressed questions concerning the non-perturbative stability of the vacuum solutions found and the presence of domain wall solutions interpolating between them. The most significant results obtained in this respect can be summarized as follows. 
\begin{itemize}
    \item We have found an explicit expression of the superpotential, extending the known formula only valid for the minimal theory \cite{Campos:2000yu, Danielsson:2016rmq}. The analytical integration of the superpotential leads to a supersymmetric domain wall solution. 
    \item By perturbatively solving the HJ equation, we have found different \emph{fake superpotentials}. Some of them ensure the non-perturbative stability - within the considered set of fields - for the solutions whose perturbative stability has been confirmed by the analysis of the mass spectra for the scalar fields. We cannot however, establish the full stability of all these vacua: perturbative or non-perturbative instabilities - expected for the non-supersymmetric vacuum \cite{Ooguri:2016pdq} - can be hidden within the truncated fields.
    \item From other branches of solutions of \eqref{eq:DW_fake_susy}, we can determine two more extremal non-supersymmetric domain walls connecting the stable vacua. One of them, according to our interpretation of the extra vector multiplets, appears to be the counterpart of a known domain wall solutions \cite{Campos:2000yu, Danielsson:2016rmq} in a case where open string effects are included.
\end{itemize}
The web of DW solutions between vacua and their properties are summarized in Figure \ref{Fig:Vacua_domain_walls}. We expect each of the domain walls discussed to correspond to an RG flow between conformal fixed points.

\section*{Acknowledgements}
We would like to thank Evangelos Afxonidis, \'Alvaro Arboleya, Adolfo Guarino, Matteo Morittu for stimulating discussions. VB thanks the Institute for Theoretical Physics at the KU Leuven for the warm hospitality while part of this project was carried out. GS thanks the Physics Department of the Oviedo University and ICTEA for the opportunity of a visiting period during the completion of this work. The work of GS is supported by ``Angelo Della Riccia'' Foundation.

\appendix

\section{Different formulations of \texorpdfstring{$D=7$}{D=7} half-maximal supergravity with 3 vector multiplets}
\label{Appendix:tHooft}
In this Appendix, we spell out the dictionary between the formulations of half-maximal supergravity theory in 7 dimensions in terms of representations of the $\mathrm{SL}(4)$ and $\mathrm{SO}(3,3)$ groups. This dictionary is entirely based on the use of the 't Hooft symbols, whose structure and explicit representation is recalled preliminarly.

\subsection{'t Hooft symbols}
The 't Hooft symbols \cite{tHooft:1976snw, Belitsky:2000ws} $[G_A]^{mn}$ are invariant tensors mapping the fundamental representation $\mathbf{6}$ of $\mathrm{SO}(3,3)$ into the anti-symmetric 2-form representation of $\mathrm{SL}(4)$. Given a vector $v^A$ in the fundamental of $\mathrm{SO}(3,3)$, the mapping is 
\begin{equation}
    v^{mn} = [G_A]^{mn} v^A \;.
\end{equation}
The 2-form irreducible representation of $\mathrm{SL}(4)$ is real, since it is linked to its conjugate $\bar{\mathbf{6}}$ via the Levi-Civita pseudotensor 
\begin{equation}
    v_{mn} = \frac{1}{2} \varepsilon_{mnpq} v^{pq}\;.
\end{equation}
Accordingly, we can define conjugate 't Hooft symbols such that
\begin{equation} \label{eq:tHooftconjugate}
    [G_A]_{mn} =  \frac{1}{2} \varepsilon_{mnpq} [G_A]^{pq} \;.
\end{equation}
They provide the inverse mapping from $\mathrm{SL}(4)$ to $\mathrm{SO}(3,3)$.

The explicit expression of the 't Hooft symbols can be deduced from the Dirac matrices in chiral representation, following the procedure in \cite{Borghese:2010ei, Dibitetto:2012ia}. The final expression that we adopt is
\begin{equation} \label{eq:tHooft}
\begin{aligned}
[G_1]^{mn} = \left(
\begin{array}{cccc}
 0 & 0 & 0 & 1  \\
 0 & 0 & 0 & 0 \\
 0 & 0 & 0 & 0 \\
 -1 & 0 & 0 & 0 
\end{array}
\right) \;,
\qquad  [G_2]^{mn} = \left(
\begin{array}{cccc}
 0 & 0 & 0 & 0 \\
 0 & 0 & 0 & 1 \\
 0 & 0 & 0 & 0 \\
 0 & -1 & 0 & 0
\end{array}
\right) \;,\\
[G_3]^{mn} = \left(
\begin{array}{cccc}
0 & 1 & 0 & 0 \\
 -1 & 0 & 0 & 0 \\
 0 & 0 & 0 & 0 \\
 0 & 0 & 0 & 0 \\
\end{array}
\right) \;,
\qquad
[G_4]^{mn} = \left(
\begin{array}{cccc}
 0 & 0 & 0 & 0  \\
 0 & 0 & -1 & 0 \\
 0 & 1 & 0 & 0 \\
 0 & 0 & 0 & 0 
\end{array}
\right) \;,\\
[G_5]^{mn} = \left(
\begin{array}{cccc}
0 & 0 & 1 & 0 \\
 0 & 0 & 0 & 0 \\
 -1 & 0 & 0 & 0 \\
 0 & 0 & 0 & 0 \\
\end{array}
\right) \;,
\qquad [G_6]^{mn} = \left(
\begin{array}{cccc}
0 & 0 & 0 & 0 \\
 0 & 0 & 0 & 0 \\
 0 & 0 & 0 & -1 \\
 0 & 0 & 1 & 0 
\end{array}
\right)\;.
\end{aligned}
\end{equation}
The corresponding conjugates are linked to these by \eqref{eq:tHooftconjugate}, as expected. The 't Hooft symbols introduced in \cite{Dibitetto:2015bia} are linked to the ones in \eqref{eq:tHooft} via a transformation belonging to $\mathrm{O}(3,3)$, acting on the vectorial indices.  

The symbols introduced in \eqref{eq:tHooft} and their conjugate satisfy the following relations \cite{Dibitetto:2012xd}:
\begin{equation} \label{eq:relations}
\begin{aligned}
& [G_A]_{mn} [G_B]^{mn} = - 2 \eta_{AB} \; , \\
& [G_A]_{mp} [G_B]^{pn} + [G_B]_{mp} [G_A]^{pn} = \delta_m^n \eta_{AB} \; , \\
& [G_{[A}]_{mp} [G_B]^{pq} [G_C]_{qr} [G_D]^{rs}[G_E]_{st} [G_{F]}]^{tn} = - \frac{1}{8} \delta_m^n \varepsilon_{ABCDEF} \; , 
\end{aligned}
\end{equation}
with the matrix $\eta_{AB}$ in light-cone coordinates
\begin{equation}
    \label{eq:etaAB}
    \eta_{AB}= \eta^{AB}= \begin{pmatrix}
        \mathbb{O}_3 & \mathbbm{I}_3 \\
        \mathbbm{I}_3 & \mathbb{O}_3 \\
    \end{pmatrix}\; .
\end{equation}

\subsection{Mapping between the \texorpdfstring{$\mathrm{SL}(4)$}{SL(4)} and \texorpdfstring{$\mathrm{SO}(3,3)$}{SO(3,3)} formulations of half-maximal supergravity}
If we consider half-maximal supergravity in 7 dimensions with the gravity multiplet plus three vector multiplets, the scalar fields of the theory parametrize the coset manifold
\begin{equation}
    \underbrace{\mathbb{R}^{+}}_{X} \times \underbrace{\frac{\mathrm{SL}(4)}{\mathrm{SO}(4)} }_{M_{mn}}\; .
\end{equation}
The coset representative $M_{mn}$ can be constructed in terms of the vielbein $\tensor{\mathcal{V}}{_{m}^{\alpha \hat{\alpha}}}$ as 
\begin{equation}
    M_{mn} = \tensor{\mathcal{V}}{_{m}^{\alpha \hat{\alpha}}} \tensor{\mathcal{V}}{_{m}^{\beta \hat{\beta}}} \tensor{\varepsilon}{_{\alpha \beta}} \tensor{\varepsilon}{_{\hat{\alpha} \hat{\beta}}}\;.
\end{equation}
Here, as well as in Subsection \ref{Subsection:7d_3_sugra}, our convention on indices is that $m,\,n$ are fundamental $\mathrm{SL}(4)$ indices, $\alpha,\, \beta$ and $\hat{\alpha},\, \hat{\beta}$ are fundamental indices of two different copies of $\mathrm{SU}(2)$, denoted with $\mathrm{SU}(2)_{R}$ and $\mathrm{SU}(2)$ and $\tensor{\varepsilon}{_{\alpha \beta}}$, $\tensor{\varepsilon}{_{\hat{\alpha} \hat{\beta}}}$ are the corresponding invariant matrices that can be used to raise and lower the indices.

The coset representative of $\mathrm{SO}(3,3)/(\mathrm{SO}(3) \times \mathrm{SO}(3) )$ is related to $M_{mn}$ through the 't Hooft symbols
\begin{equation}
    M_{AB}= \frac{1}{2} M_{mn}\, M_{pq}\, [\tensor{G}{_A}]^{mp}\, [\tensor{G}{_B}]^{nq}\;.
\end{equation}

The possible deformations of the theory are parametrized by $\{\theta, \, Q_{mn}, \, \tilde{Q}^{mn},\, \xi_{mn}\}$. Among them, the gauge algebra generators can be rewritten as
\begin{equation}
    \tensor{\big( X_{mn}\big)}{_{pq}^{rs}} = \frac{1}{2} \,\tensor{\delta}{_{[m}^{[r}}\, Q_{n][p} \,\tensor{\delta}{_{q]}^{s]}} - \frac{1}{4} \,\varepsilon_{tmn[p} \,\big(\tilde{Q} + \xi \big)^{t [r}\, \tensor{\delta}{_{q]}^{s]}}\;.
\end{equation}
With this definition, it is possible to embed all the deformations of the theory into $\mathrm{SO}(3,3)$ representations with the following mapping
\begin{equation}
\label{eq:7d_SO3,3_embedding_f}
    f_{ABC} = \tensor{\big( X_{mn}\big)}{_{pq}^{rs}}\, [\tensor{G}{_A}]^{mn}\, [\tensor{G}{_B}]^{pq}\, [\tensor{G}{_C}]_{rs}\;,
\end{equation}
\begin{equation}
\label{eq:7d_SO3,3_embedding_xi}
    \xi_A = \xi_{mn} \, [\tensor{G}{_A}]^{mn} \;,
\end{equation}
while the parameter $\theta$ is a scalar in both representations.

The expression of the scalar potential in the $\mathrm{SL}(4)$ formalism is
\begin{equation}
    \begin{split}
        V=& \,\frac{1}{64} \bigg( \theta^2 \, X^8 + \frac{1}{4} Q_{mn} \,Q_{pq}\, X^{-2} \big( 2 M^{mp} M^{nq} - M^{mn} M^{pq} \big) +\\
        & + \frac{3}{2} \xi_{mn}\, \xi_{pq}\, X^{-2}\, M^{mp} M^{nq} + \frac{1}{4} \, \tilde{Q}^{mn} \tilde{Q}^{pq} \, X^{-2} \big( 2 M_{mp} M_{nq} - M_{mn} M_{pq} \big)+\\
        &- \theta \, X^3 \big(Q_{mn} M^{mn} - \tilde{Q}^{mn} M_{mn} \big) + Q_{mn} \tilde{Q}^{mn} X^{-2} \bigg) \;, 
    \end{split}
\end{equation}
where the matrix $M^{mn}$ is the inverse of the coset representative $M_{mn}$. In the $\mathrm{SO}(3,3)$ formalism, and setting $\xi_A=0$, it becomes
\begin{equation} \label{eq:7d_SO3,3_Potential}
\begin{split}
    V=& \;\frac{1}{64} \bigg[ \frac{X^{2}}{2} f_{ABC} f_{DEF} \Big( \frac{1}{3} M^{AD} M^{BE} M^{CF} + \Big( \frac{2}{3} \eta^{AD} - M^{AD}\Big) \eta^{BE} \eta^{CF} \Big) +  \\
    &+ \theta^2 X^{-8} -\frac{2 \sqrt{2}}{3} \theta X^{-3}  f_{ABC} M^{ABC} \bigg]\;,
\end{split} 
\end{equation}

\section{Massive IIA supergravity in the democratic formulation}
\label{Appendix:IIA_democratic}
In this Appendix, we provide some details about type IIA supergravity in $10$ dimensions in presence of Romans' mass \cite{Romans:1985tz}, usually referred to as \emph{massive type IIA supergravity}. This theory admits a democratic formulation \cite{Bergshoeff:2001pv}, which we adopt in Section \ref{Section:7d_10d_Reduction}. Its bosonic fields consist of the NS-NS fields $\{G, B_{(2)}, \Phi\}$, plus the Ramond-Ramond degrees of freedom, which are doubled with respect to the standard formulation $\{C_{2p + 1}\}_{p=0, 1, 2, 3, 4}$. The equations of motion for the bosonic sector can be obtained by varying the following pseudoaction:
\begin{equation} \label{eq:pseudoaction}
S_{\textup{II A}}^{\textup{democratic}} = \frac{1}{2 \kappa_{10}^2} \int \mathrm{d}^{10} x \ \sqrt{-G} \Bigl (e^{-2 \Phi} \bigl (\mathcal{R} + 4 (\partial \Phi)^2 - \frac{1}{12} |H_{(3)}|^2   \bigr ) - \frac{1}{4} \sum_{p=0}^5 \frac{|F_{(2 p)}|^2}{(2 p)!} \Bigr) \, ,
\end{equation} 
where we have defined the field strengths 
\begin{equation}
H_{(3)} = \dd B_{2} \,, \qquad 
    \quad F= \dd C - \dd B \wedge C + F_{(0)} \, e^{B} \;.\label{eq:bosonic_field_strengths}
\end{equation}
The following duality relations are needed in order to restore the correct counting of the degrees of freedom in the R-R sector
\begin{equation} \label{eq:duality}
     F_{(0)} = \star F_{(10)} \;, \qquad  F_{(2)} = - \star F_{(8)} \;, \qquad  F_{(4)} = \star F_{(6)} \;, \qquad H_{(7)} = e^{-2 \Phi} \star H_{(3)} \, .
\end{equation}
By definition, the field strengths \eqref{eq:bosonic_field_strengths} obey the following modified Bianchi identities
\begin{align}
     \dd H_{(3)} = 0 \quad, & \qquad  \dd H_{(7)} -\frac{1}{2} \sum_{p=0}^4 F_{(2p)} \wedge \star F_{(2p+2)} = 0 \quad, \label{eq:modifiedBianchiH}\\
    \dd F_{(0)} = 0 \quad, & \qquad  \dd F_{(2)} - H_{(3)} \wedge F_{(0)}  = 0 \quad,  \label{eq:modifiedBianchiF0F2} \\
    \dd F_{(4)} - H_{(3)} \wedge F_{(2)} = 0 \quad, & \qquad  \dd F_{(6)} - H_{(3)} \wedge F_{(4)} = 0 \quad, \label{eq:modifiedBianchiF4F6}\\
    \dd F_{(8)} - H_{(3)} \wedge F_{(6)} = 0 \quad, & \qquad  \dd F_{(10)} = 0 \;.
    \label{eq:modifiedBianchiF8F10}
\end{align}
Making use of the duality relations \eqref{eq:duality}, we can map the Bianchi identities \eqref{eq:modifiedBianchiH}, \eqref{eq:modifiedBianchiF0F2}, \eqref{eq:modifiedBianchiF4F6}, \eqref{eq:modifiedBianchiF8F10} to the equations of motion for the dual fields, either in the NS-NS or the R-R sector
\begin{align}
     \dd \bigl (e^{2 \Phi}  \star H_{(7)} \bigr ) = 0 \quad, & \qquad \dd \bigl ( e^{- 2 \Phi}  \star H_{(3)} \bigr ) - \frac{1}{2} \sum_{p=0}^4 F_{(2p)} \wedge \star F_{(2p+2)} = 0 \quad, \\
    \dd \star F_{(10)} = 0 \quad, & \qquad \dd \star F_{(8)} + H_{(3)} \wedge \star F_{(10)}  = 0 \quad,  \\
    \dd \star F_{(6)} + H_{(3)} \wedge \star F_{(8)} = 0 \quad, & \qquad \dd \star F_{(4)} + H_{(3)} \wedge \star F_{(6)} = 0 \quad, \\
    \dd \star F_{(2)} + H_{(3)} \wedge \star F_{(4)} = 0 \quad, & \qquad \dd \star F_{(0)} = 0 \quad. 
\end{align}
In particular, the equations of motion in the R-R sector can be formally written as
\begin{equation}
\dd \star F_{(2p)} + H_{(3)} \wedge \star F_{(2p + 2)} = 0 \;.
\end{equation}
As for the remaining fields, the dilaton has the following equation of motion
\begin{equation}
\label{eq:Dilaton_eom_base}
\nabla_M \nabla^M \Phi - \nabla_M \Phi \nabla^M \Phi + \frac{1}{4} \mathcal{R} - \frac{1}{8 \times 3!} |H_{(3)}|^2 = 0 \quad,
\end{equation}
while the Einstein's equations are
\begin{multline}\label{eq:Einstein_eom_base}
e^{- 2 \Phi} \big(\mathcal{R}_{MN} + 2 \nabla_M \nabla_N \Phi - \frac{1}{4} H_{MPQ}{H_N}^{PQ} \big) - \frac{1}{2} (F_{(2)}^2)_{MN} - \frac{1}{2 \times 3!} (F_{(4)}^2)_{MN} + \\  +  \frac{1}{4} G_{MN} \big(|F_{0}|^2 + \frac{1}{2!} |F_{(2)}|^2 + \frac{1}{4!}|F_{(4)}|^2 \big) =0\quad.
\end{multline}

\section{Non-abelian D-brane effective action}
\label{Appendix:NonAbelian_DBI_WZ}
Due to the non-Abelianity of the fields involved, the effective action of a stack of coincident D$p$-branes gets modifications with respect to the action for single branes. 
In this Section, we describe the key features of this action \cite{Myers:1999ps, Martucci:2005rb}, following the notation adopted in \cite{Choi:2018fqw}, \cite{Balaguer:2023jei}. The action splits into two parts: the Dirac-Born-Infeld (DBI) and Wess-Zumino (WZ) actions. Written in the string frame, they take the form:
\begin{equation}
\label{eq:DBI}
    S^{\textup{DBI}}_{\textrm{D}p}= -T_{\textrm{D}p} \int_{\textup{WV}(\textrm{D}p)} \dd^{p+1} \xi \; \Tr \bigg( e^{-\hat{\Phi}} \sqrt{-\det(\mathbb{M}_{\mu \nu}) \det(\tensor{\mathbb{Q}}{^i_j})} \; \bigg) \quad,
\end{equation}
\begin{equation}
\label{eq:WZ}
    S^{\textup{WZ}}_{\textrm{D}p}= \mu_{\textrm{D}p} \int_{\textup{WV}(\textrm{D}p)}  \Tr \bigg\{ \mathrm{P} \bigg[ e^{i \lambda \iota_Y \iota_Y} \Big(\boldsymbol{\hat{C}} \wedge e^{\hat{B}_{(2)}} \Big) \wedge e^{\lambda \mathcal{F}}  \bigg] \bigg\} \quad,
\end{equation}
where we limit our attention to the bosonic part of the action.
The matrices $\mathbb{M}_{\mu \nu}$ and $\tensor{\mathbb{Q}}{^i_j}$ appearing in the DBI action are
\begin{equation} \label{eq:MMN}
    \mathbb{M}_{\mu \nu}= \mathrm{P} \Big[ \hat{E}_{\mu \nu} + \hat{E}_{\mu i} (\mathbb{Q}^{-1} - \delta )^{ij} \hat{E}_{j \nu} \Big] + \lambda \mathcal{F}_{\mu \nu} \quad,
\end{equation}
\begin{equation} \label{eq:Qij}
    \tensor{\mathbb{Q}}{^i_j}=\tensor{\delta}{^i_j}+i \lambda [Y^i,Y^k] \hat{E}_{kj} \quad,
\end{equation}
where
\begin{equation} \label{eq:EMN}
    \hat{E}_{MN}=\hat{g}_{MN}+\hat{B}_{MN} \quad .
\end{equation}
In the expression \eqref{eq:WZ} for the WZ action, $\boldsymbol{C}$ stands for the polyform corresponding to the sum of all the $C_{(p)}$ fields, $\iota_{Y}$ is the interior product by the $Y^i$ vector and $\mathcal{F}$ is the non-Abelian field-strength of the gauge field $\mathcal{A}$ of the super Yang-Mills theory along the branes:
\begin{equation} \label{eq:Nonab_FieldStrength}
    \mathcal{F} = \mathrm{d} \mathcal{A} + i \mathcal{A} \wedge \mathcal{A} \;.
\end{equation}

The position of the $Dp$-branes in the transverse space is encoded in $y^i= \lambda Y^{Ii} t_{I}$, where $t_I$ are the generators of the Yang-Mills gauge algebra, such that
\begin{equation} \label{eq:GeneratorRelations}
    [t_I, t_J] = -i {g_{IJ}}^K t_K \quad, \quad \mathrm{Tr} [t_I t_J] = N_{Dp}\,\delta_{IJ} \quad.
\end{equation}
$N_{Dp}$ is the number of branes, not to be confused with the dimension of the gauge algebra $\mathfrak{N}$ \cite{Gimon:1996rq, Giveon:1998sr, Bergman:2001rp, Balaguer:2023jei} (the two quantities only coincide in the Abelian case). The constant $\lambda$ is related to the string length according to $\lambda= 2 \pi {\ell_S}^2$. 

$\mathrm{P} [-] $ is the pullback of the bulk fields over the $D6$-brane worldvolume. For instance, the pullback of $\hat{E}_{\mu \nu}$ is
\begin{equation} \label{eq:PullBack}
    \mathrm{P}[\hat{E}_{\mu \nu}]= \hat{E}_{\mu \nu}+ \lambda D_{\mu} Y^i \hat{E}_{i\mu} + \lambda D_{\nu} Y^i \hat{E}_{\mu i} +\lambda^2  D_{\mu} Y^i D_{\nu} Y^j \hat{E}_{ij}  \;.
\end{equation}
Here, the covariant derivative is 
\begin{equation} \label{eq:Covariant_Derivative}
    D_{\mu} Y^i \equiv \partial_{\mu} Y^i - i [\mathcal{A}_{\mu}, Y^i] \;.
\end{equation}
All the hatted fields are evaluated at the position of the $D6$-branes with a Taylor expansion around $y^i=0$ (static gauge):
\begin{equation}
    \hat{\Phi}(x^\mu, y^i)= \sum_{k=0}^{\infty} \frac{\lambda^k}{k!} Y^{i_1}\cdots Y^{i_k} \partial_{i_1} \cdots \partial_{i_k} \Phi(x^\mu, y^i) \Big|_{y^i=0} \;.
\end{equation}

The contributions to the bosonic effective action for the orientifold plane in the string frame, instead, are given by
\begin{equation} \label{eq:OrientifoldDBIAction}
S^{\mathrm{DBI}}_{\mathrm{O}p} = - T_{\mathrm{O}6} \int \mathrm{d}^7 \xi \ e^{- \Phi} \sqrt{- \mathrm{det} (g_{\mu \nu})} \quad,
\end{equation}
\begin{equation} \label{eq:WZOrientifold}
S^{\mathrm{WZ}}_{\mathrm{O}p} = \mu_{\mathrm{O}6} \int_{WV(\mathrm{O}p)} \hat{C}_{(p+1)} \quad.\end{equation}
Compared to \eqref{eq:DBI} and \eqref{eq:WZ}, no other terms appear, since orientifold planes are not dynamical objects, at least perturbatively. 

\section{Local branches of solutions to the HJ equation}
\label{Appendix:HJ_branches}

In this Appendix, we complement the discussion of Subsection \ref{Subsec:FakeSusy_solutions} with  additional details about the perturbative solutions to the HJ equation around the critical points of half-maximal 7D supergravity listed in Table \ref{Tab:7d_Solutions_FixedFluxes}. 

When we solve perturbatively around the supersymmetric vacuum $\mathbf{1'}$, we obtain
\begin{itemize}
    \item at zero order $f^{(0)} =\frac{5}{32} q $,
    \item the eight branches of solutions for the superpotential are associated to eight possible solutions to the Hamilton-Jacobi equation at order two, which correspond to the following possibilities for the matrix of second-order coefficients:
    \begin{equation}
        f^{(2)} = \text{diag} \bigg(\frac{5 (3 \pm 1)}{32} q, \,\frac{3 (3 \pm 7 )}{32} q,\, \frac{3(3\pm 7)}{64} q \bigg) \;,
    \end{equation}
    where the variables in the expansion of the superpotential are the scalar fields of \eqref{eq:DW_SO3_scalars}, i.\,e. $\{ X, T,Y\}$.
\end{itemize}

The analytic solutions correspond to the following branches
\begin{equation}
    \begin{split}
        f^{\mathbf{1'} a} \;: & \quad f^{(2)}=\text{diag} \bigg(\frac{5}{8} q, \,-\frac{3}{8} q,\, -\frac{3}{16} q \bigg) \;,\\
        f^{\mathbf{1'} b} \;: & \quad f^{(2)}= \text{diag} \bigg(\frac{5}{8} q, \,\frac{15}{16} q,\, \frac{15}{32} q \bigg)\;.
    \end{split}
\end{equation}

Concerning the solutions around the non-supersymmetric point $\mathbf{3'}$, we have
\begin{itemize}
    \item at zero order $f^{(0)} = \frac{5}{2^{1/5} \sqrt{3}\cdot 16} q$,
    \item at order two, the eight branches are characterized by
    \begin{equation}
        f^{(2)} = \text{diag} \bigg(\frac{ 5\cdot 2^{1/5}(\sqrt{3} \pm \sqrt{7})}{16 } q, \,\frac{ 9 \,(1 \pm \sqrt{5} )}{16 \cdot 2^{1/5} \sqrt{3}} q,\, \frac{3 \sqrt{3} (1 \pm \sqrt{5})}{32 \cdot 2^{1/5}} q \bigg)\;.
    \end{equation}
\end{itemize}
In this case, the analytic solutions correspond to the following branches
\begin{equation}
    \begin{split}
        f^{\mathbf{3'} a} \;: & \quad f^{(2)}=\text{diag} \bigg(\frac{ 5\cdot 2^{1/5}(\sqrt{3} - \sqrt{7})}{16 } q, \,\frac{ 9 \,(1 - \sqrt{5} )}{16 \cdot 2^{1/5} \sqrt{3}} q,\, \frac{3 \sqrt{3} (1 - \sqrt{5})}{32 \cdot 2^{1/5}} q \bigg) \;,\\
        f^{\mathbf{3'} b} \;: & \quad f^{(2)}=\text{diag} \bigg(\frac{ 5\cdot 2^{1/5}(\sqrt{3} + \sqrt{7})}{16 } q, \,\frac{ 9 \,(1 + \sqrt{5} )}{16 \cdot 2^{1/5} \sqrt{3}} q,\, \frac{3 \sqrt{3} (1 + \sqrt{5})}{32 \cdot 2^{1/5}} q \bigg) \;.
    \end{split}
\end{equation}

The solutions around the supersymmetric point $\mathbf{1}$ feature
\begin{itemize}
    \item at zero order $f^{(0)} = \frac{5}{16 \cdot 2^{3/5} \cdot 3^{2/5}} q$,
    \item at order two
    \begin{equation}
        f^{(2)} = \text{diag} \bigg(\frac{5 \,(3 \pm 1)}{32}\Big(\frac{2}{3}\Big)^{3/5} q, \,\frac{  9 \,(3 \pm 7 )}{32 \cdot 2^{3/5} \cdot 3^{2/5}} q,\, \frac{3 \pm 1}{32}\Big(\frac{2}{3}\Big)^{2/5} q\bigg)\;.
    \end{equation}
\end{itemize}
The analytic solutions correspond to the following branches
\begin{equation}
    \begin{split}
        f^{\mathbf{1} a} \;: & \quad f^{(2)}= \text{diag} \bigg(\frac{5 }{8}\Big(\frac{2}{3}\Big)^{3/5} q, \,-\frac{  9 }{8 \cdot 2^{3/5} \cdot 3^{2/5}}q,\, \frac{1}{8}\Big(\frac{2}{3}\Big)^{2/5} q\bigg)\;,\\
        f^{\mathbf{1} b} \;: & \quad f^{(2)}= \text{diag} \bigg(\frac{5}{8}\Big(\frac{2}{3}\Big)^{3/5} q, \,\frac{  45}{16 \cdot 2^{3/5} \cdot 3^{2/5}} q,\, \frac{1}{8}\Big(\frac{2}{3}\Big)^{2/5} q\bigg)\;.
    \end{split}
\end{equation}

For completeness, we also report the results of the computation around the vacuum $\mathbf{3}$, which is an unstable solution:
\begin{itemize}
    \item at zero order $f^{(0)} = \frac{5 \cdot 2^{1/5}}{16 \cdot 3^{9/10}} q$,
    \item at order two, solving the equations with respect to the matrix of second order derivatives gives eight complex-value solutions:
    \begin{equation}
        f^{(2)} = \text{diag} \bigg(-\frac{ 5\cdot 2^{4/5}(3 \pm \sqrt{21})}{16 \cdot 3^{11/10}} q, \,-\frac{27 \, (1 \pm \sqrt{5} )}{16 \cdot 2^{4/5}\cdot 3^{9/10}} q,\, \frac{2^{1/5} (-3 \pm i \sqrt{3})}{16 \cdot 3^{9/10}} q \bigg)\;.
    \end{equation}
    This is expected, due to the violation of the BF bound, as can be seen from equations \eqref{eq:DW_f2_coefficients} and \eqref{eq:DW_f2_coeff_BF_bound}. In particular, the instability comes from the scalar field $Y$ and the non-real values for the derivatives are found in the corresponding direction.
\end{itemize}

\bibliographystyle{jhep}
\bibliography{bibliography.bib}

\end{document}